\documentclass{JFM-FLM_Au}

\lefttitle{Raagvendra Singh et al.}
\righttitle{Journal of Fluid Mechanics}

 \title{Sub-filter-scale stress and heat flux in hypersonic Couette flow and a flat-plate boundary layer}

\author{Raagvendra Singh\aff{1},
  Sabbir Ahmed\aff{2}, Lian Duan\aff{2}, Carlo Scalo\aff{1}, \and Takahiko Toki\aff{1}}

\affiliation{\aff{1}Department of Mechanical Engineering, Purdue University, IN 47907, USA
\aff{2}Department of Mechanical and Aerospace Engineering, The Ohio State University, OH 43210, USA}

\corresau{Raagvendra Singh, sing1217@purdue.edu}

\begin{document}
\pagestyle{plain}

\maketitle

\begin{abstract}

A comparative study of sub-filter-scale (SFS) quantities in hypersonic Couette flow and flat-plate boundary layer turbulence under strongly cooled-wall conditions is carried out using filtered direct numerical simulation (DNS) data to characterize the differences in SFS dynamics between internal and external wall-bounded turbulence. The two flows exhibit similar near-wall behaviour in the mean velocity profiles and SFS stress components; however, marked differences emerge in the outer layer and thermal structure. In the Couette flow, turbulence persists throughout the domain, whereas in the flat-plate boundary layer the turbulence intensity decreases towards the boundary-layer edge. The two configurations also exhibit different thermal fields, leading to different temperature–velocity relations. The SFS behaviour shows strong sensitivity to the filtering operation: the top-hat filter yields larger SFS magnitudes, whereas the sharp spectral filter preserves a distinct spectral separation near the cutoff, leading to characteristic SFS features. Decomposition of the SFS terms shows that the residual component provides the dominant contribution to the mean SFS shear stress and heat flux. Differences between the two flows are primarily associated with the interaction terms between the resolved and residual components, which contribute with opposite signs in the two configurations, consistent with their different thermal structures. The different thermal structures also lead to distinct SFS heat-flux behaviour, including a sign change in the flat-plate boundary layer. Instantaneous SFS fields further reveal that local extrema arise from the combined contributions of multiple decomposed terms. Although these terms exhibit large local fluctuations, their net contributions to the averaged SFS quantities are relatively small. The trace of the SFS stress exhibits local negative values, reflecting local energy redistribution across the filter scale rather than a loss of physical energy.

\end{abstract}


\section{Introduction}

\subsection{Motivation}

Wall-bounded supersonic and hypersonic turbulent flows play a critical role in the aerodynamic heating and thermal design of high-speed vehicles. Accurate prediction of compressible turbulent flows is therefore essential, and these flows have consequently been extensively investigated using high-fidelity approaches such as direct numerical simulations (DNS) and large eddy simulations (LES) \citep{martin2007, duan2010direct, duan2011a, duan2011b, zhang2018, huang2022, yu2021compressibility}.
In LES, the effects of unresolved small scales are represented through sub-filter-scale (SFS) terms, which play a central role in the transfer of momentum and energy across scales.

Wall-bounded turbulent flows can be broadly categorised into internal and external configurations. Internal flows, such as channel and Couette flows, 
are confined by rigid boundaries, whereas external flows, such as boundary layers over flat plates, 
develop in the presence of a freestream. As a result, turbulence in internal flows extends across the flow domain, while in external flows, it is confined within a boundary layer. From a computational perspective, internal flows are often more amenable to high-fidelity simulations, as periodic boundary conditions can be imposed in the homogeneous directions. In contrast, external flows are spatially developing and typically require larger computational domains, leading to significantly higher computational cost for DNS and LES.

Consequently, internal flows have frequently been used as canonical configurations for investigating turbulent dynamics. However, it is still uncertain how far the insights gained from hypersonic internal flows can be directly transferred to external flows, especially regarding the SFS stress and heat-flux terms that are important for LES. This motivates a detailed comparison of these terms between the two configurations in order to assess their similarities and differences and to establish their respective characteristics in compressible turbulent flows, which lays the foundation for developing physics based SFS models for LES of hypersonic flows.

\subsection{Background and objectives}

DNS have provided detailed insight into hypersonic turbulent flows. Early studies examined turbulent boundary layers and highlighted the influence of wall temperature, Mach number, and compressibility on turbulence statistics \citep{martin2007, duan2010direct, duan2011a, duan2011b, huang2022}. DNS has also been applied to internal compressible flows, including channels and Couette flows, to investigate the role of thermodynamic variations on velocity and temperature fluctuations \citep{yu2021compressibility, ChenS_JFM_2021}.

The high computational cost of DNS, however, limits its applicability to realistic configurations. As a result, mid- to low-fidelity approaches such as Reynolds-averaged Navier--Stokes (RANS) and LES are commonly employed for practical predictions. While RANS models can provide estimates of mean flow quantities, they often fail to capture unsteady phenomena and complex interactions, particularly in high-speed regimes involving shock--boundary-layer interactions. In contrast, LES provides a more suitable framework by resolving the large-scale turbulent structures while modelling the unresolved small-scale motions. 
The effects of the unresolved small-scale motions, captured by the SFS terms, is crucial for mediating the transfer of momentum and energy between different scales.

A number of studies have therefore examined SFS behaviour in compressible turbulent flows. For example, \citet{helm2022} quantified the contribution of SFS terms to turbulent kinetic energy, while \citet{xu2021b} analysed the influence of wall temperature on kinetic energy transfer in hypersonic boundary layers.
\citet{pati2024comparison} highlighted how strongly near-wall quantities depend on SFS modelling by comparing DNS and LES of hypersonic boundary layers.

\citet{toki2024sub} investigated SFS shear stress in hypersonic Couette flow and reported a shear-stress deficit associated with compressibility effects. Recent studies have also explored unified formulations of shock-capturing and SFS modelling for compressible turbulent flows \citep{SousaQSV_JCP_2022, ruggeri2025sub}.

Despite these efforts, most investigations have treated internal and external flows separately, and systematic comparisons of SFS dynamics between canonical compressible wall-bounded flows remain limited, to the best of the authors' knowledge, particularly under hypersonic cold-wall conditions. Differences between Couette flow and flat-plate boundary layers in terms of SFS shear stress, stress trace, and heat flux therefore remain unclear.

The objective of the present study is therefore to analyse and compare the SFS dynamics of hypersonic Couette flow and flat-plate boundary layers, and to identify their similarities and differences in the high-Mach-number and cold-wall regime representative of practical high-speed applications. The SFS terms are derived directly by filtering the DNS data, allowing an exact examination that does not rely on modelling assumptions. The study focuses in particular on the behaviour of SFS shear stress, stress trace, and heat flux in hypersonic turbulent flows.

\subsection{Paper outline}
The remainder of the paper is organized as follows. The computational setup and flow conditions are summarized in \S~\ref{sec:comp_setup}. The numerical methods employed for the present study, along with the governing equations used for LES, are described in \S~\ref{sec:Numerical method}. Time-averaged statistics, with particular emphasis on the similarities and differences between the Couette and flat-plate flows, are analysed in \S~\ref{sec:time-average}. Specifically, mean flow profiles are presented in \S~\ref{Mean flow profiles}, fluctuation correlations are discussed in \S~\ref{second-order statistics}, and the energy spectra are analysed in \S~\ref{sec:spectra}. In \S~\ref{SFS analysis}, the SFS behaviour of the Couette and flat-plate flows is compared. The effects of filter type, filter width, and the decomposition of the SFS terms are examined in \S\S~\ref{Effect of filter type}--\ref{Decomposition of SFS terms}, while the instantaneous flow fields and SFS quantities are investigated in \S~\ref{Turbulent Structural Analysis}. The main findings of the study are summarized in \S~\ref{Conclusion}.

\section{Computational Setup}\label{sec:comp_setup}

\begin{figure}
\centering
\begin{tabular}
[c]{c}%
\includegraphics[width=0.8\textwidth]{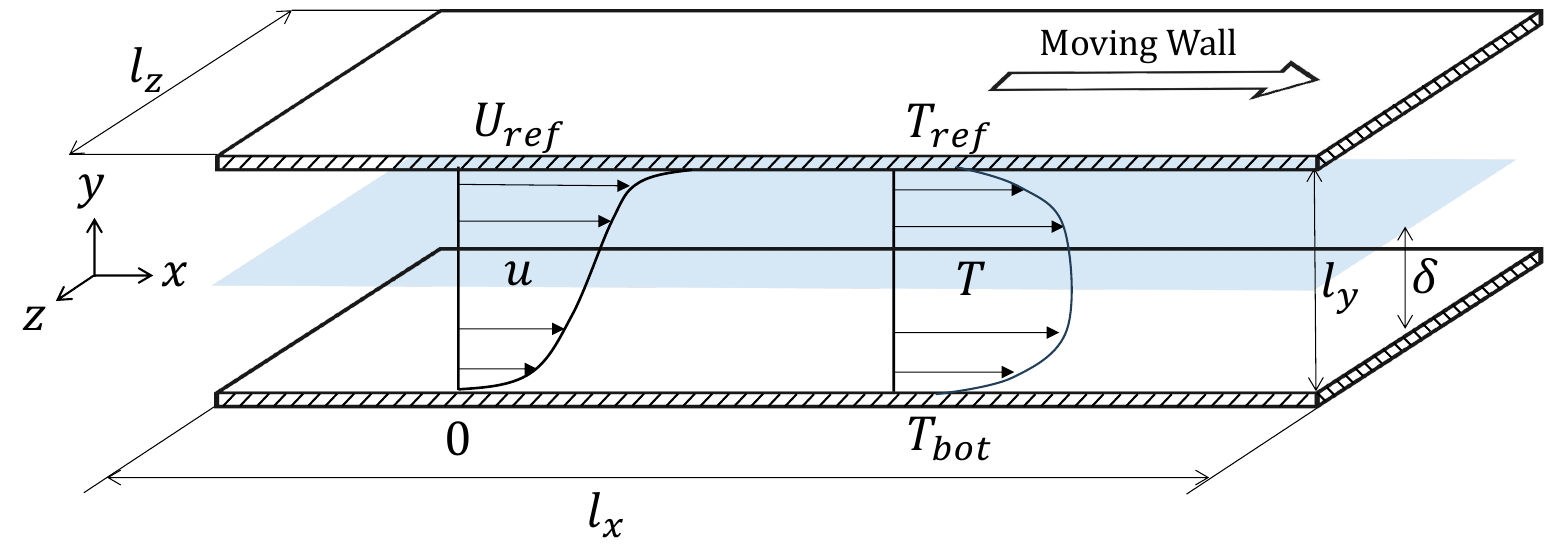}
\put(-328,100){(a)}
\\
\\
\includegraphics[width=0.9\textwidth]{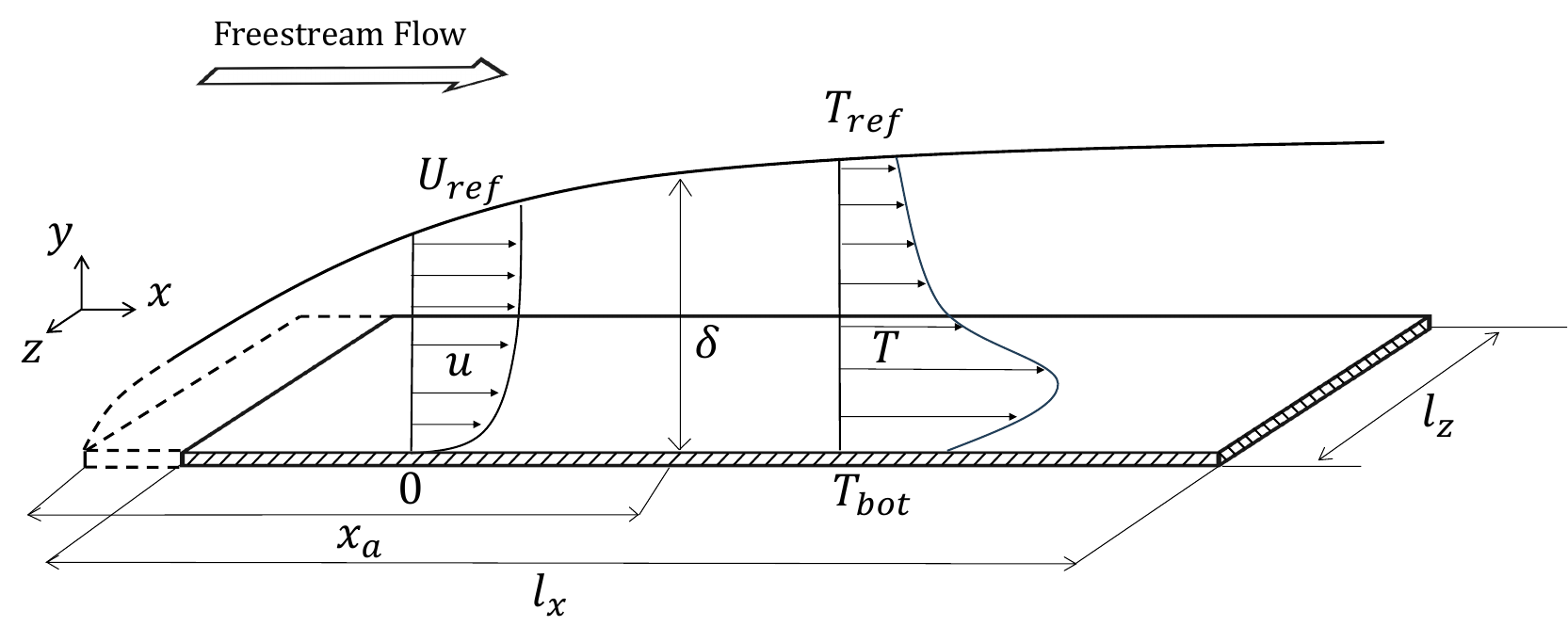}
\put(-345,100){(b)}
\end{tabular}
\caption{Sketch of the flow configuration for (a) the Couette flow and (b) flat-plate boundary layer.} 
\label{fig:sketch}
\end{figure}

\begin{figure}
\centering
\begin{tabular}
[c]{c}%
\includegraphics[width=0.6\textwidth]{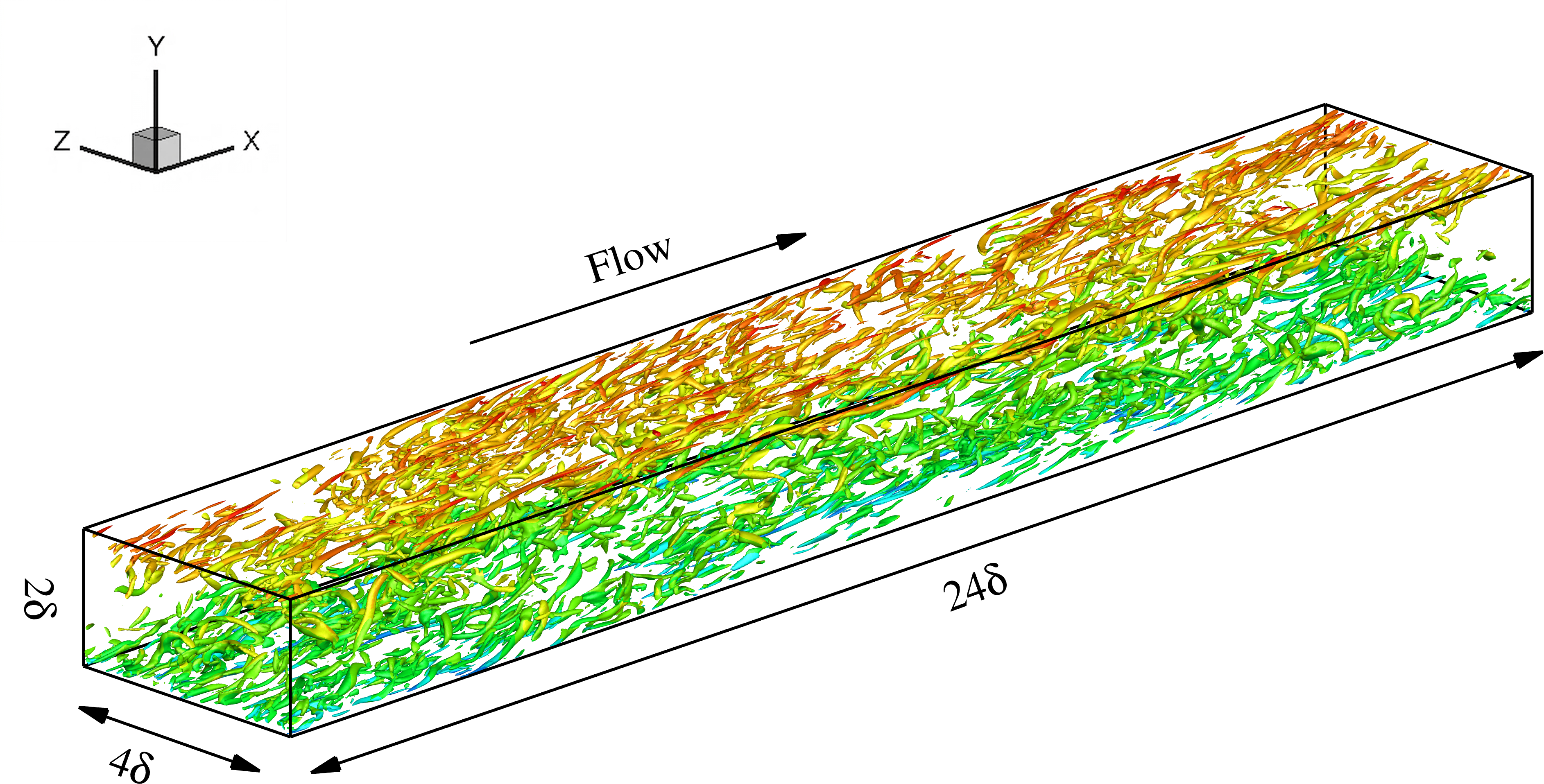}
\put(-310,100){(a)}
\\
\\
\includegraphics[width=0.9\textwidth]{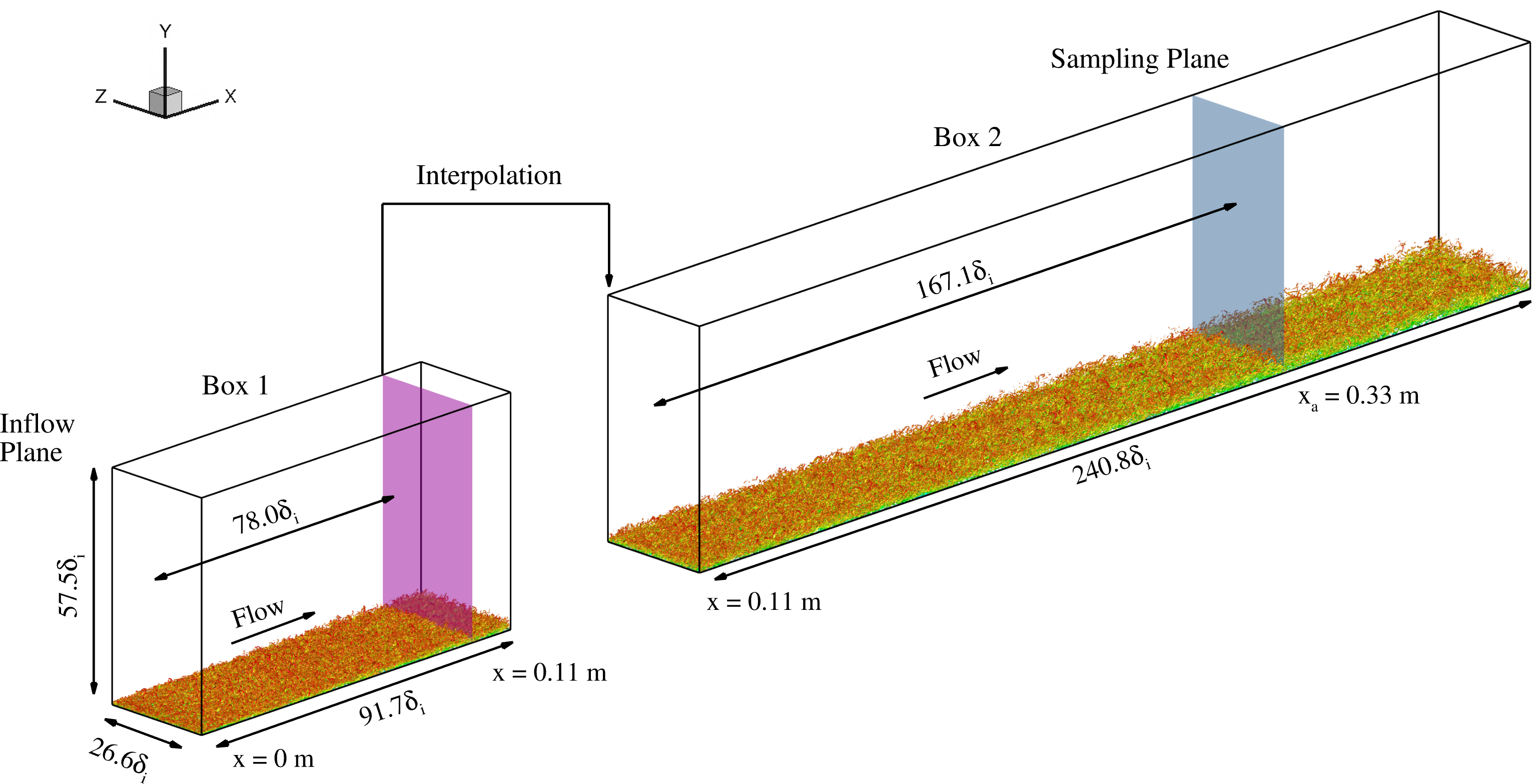}
\put(-345,100){(b)}
\end{tabular}
\caption{ Computational domain setup for DNS cases of hypersonic (a) Coutte flow and, (b) turbulent boundary layers, showing an instantaneous flow visualized by an iso-surface of Q-criterion. The iso-surface is colored by the streamwise velocity component, ranging from $0$ to $U_\text{ref}$ (blue to red).
For turbulent Couette flow, $\delta$ represents half channel height, with $\delta = 10.0$ mm. For turbulent boundary layer, $\delta_i$ represents the boundary layer thickness at the inflow plane, with $\delta_i = 1.33$ mm.} 
\label{fig:sketch_alt}
\end{figure}

\begin{table}
\centering
\begin{tabular}[h]{l c c c c c c c c c}
Case & Re${_\tau}$ & $T_{\text{bot}}$ & $T_{\text{ref}}$  &  $U_{\text{ref}}$ & $M_{\text{ref}}$  &  $\rho_{\text{ref}}$  & $T_{\text{bot}}/T_r$  & $\delta$ & $N_x \times N_y \times N_z$ \\
         &  & $(\mathrm{K})$ & $(\mathrm{K})$ &  $(\mathrm{m s^{-1}} )$ &   & $(\mathrm{kgm^{-3}})$ & & (mm) &\\  \\
Couette     &713 & 300 & 224 & 2406 & 8.0  & 0.013 & 0.112 & 10.0 &$512\times 192 \times 256$\\
Flat plate  &713 & 97.5 & 55.2 & 869.1 & 5.84 & 0.044 & 0.25 & 6.3 & $6300{\times}623{\times}1200$
\end{tabular}
\caption{List of compared cases and flow conditions. The subscript `$\text{ref}$' indicates reference quantities defined for comparison between the two flows. For the Couette flow, $T_{\text{ref}}$ and $U_{\text{ref}}$ are taken as the top-wall temperature and velocity, $\rho_{\text{ref}}$ is the bulk density, and $\delta$ is the channel half-width, whereas for the flat-plate boundary layer they correspond to the free-stream temperature, velocity, and density, and the boundary-layer thickness at the streamwise location where $Re_{\tau}$ is evaluated. The reference Mach number $M_{\text{ref}}$ is defined based on $U_{\text{ref}}$ and $T_{\text{ref}}$.}
\label{tbl:conditions}
\end{table}

The present analysis is based on the DNS datasets of \cite{toki2024sub} and \cite{ahmed_aiaa_2024} for the Couette flow and the flat-plate boundary layer, respectively. The corresponding flow configurations are shown in figure \ref{fig:sketch}.

In the Couette flow, the top wall moves at a constant velocity, while the bottom wall remains stationary. The distance between the two walls, denoted by \( l_y \), is equal to \( 2\delta \), where \( \delta \) is the channel half-width. The domain length in the streamwise direction is \( 24\delta \), and that in the spanwise direction is \( 4\delta \). No-slip, isothermal boundary conditions are imposed on both walls. Periodic boundary conditions are applied in the streamwise (\( x \)) and spanwise (\( z \)) directions. The top wall Mach number is 8. The detailed flow parameters are provided in Table~\ref{tbl:conditions}. The top wall acts as a surrogate for the edge of the boundary layer and is maintained at a temperature \( T_{\text{ref}} = 224 \, \mathrm{K} \), while the bottom wall is set at \( T_{\text{bot}} = 300 \, \mathrm{K} \). The bulk pressure \( p_{\text{bulk}} \), computed by volume-averaging the pressure over the entire computational domain, is 2950 Pa. The top wall velocity is \( U_{\text{ref}} = 2406 \, \mathrm{ms^{-1}} \), and the ratio of the bottom wall temperature to the recovery temperature, \( T_{\text{bot}} / T_r \), is 0.112.
The recovery temperature $T_r$ is defined as 

\begin{equation}
    T_r \equiv T_{\text{ref}} \left(1+Pr^{1/3}\frac{\gamma - 1}{2}{M_{\text{ref}}}^2 \right)
    \label{eq:recovery}
\end{equation}
where $Pr$ is Prandtl number ($Pr = 0.72$), $\gamma$ is specific heat ratio ($\gamma = 1.4$). In table \ref{tbl:conditions}, $Re_{\tau} \equiv u_{\tau} \delta \rho _{\text{bot}} / \mu _{\text{bot}}$ is the friction Reynolds number where $\rho _{\text{bot}}$ is the density, $\mu _{\text{bot}}$ is the viscosity and $u_{\tau}\equiv \sqrt{\tau _{\text{bot}}/\rho_{\text{bot}}}$ is the friction velocity at the bottom wall. Here, $\tau_{\text{bot}}$ is the wall shear stress at the bottom wall.

The simulation of a flat plate boundary layer is performed at a freestream Mach number of $M_\text{ref}=5.84$. Table~\ref{tbl:conditions} summarises the freestream and wall temperature conditions for the flat plate DNS. The freestream velocity and temperature, referred to as $U_\text{ref}$ and $T_\text{ref}$ respectively to be consistent with those of the Couette flow, are \( U_{\text{ref}} = 869.1 \, \mathrm{ms^{-1}} \) and \( T_{\text{ref}} = 55.2 \, \mathrm{K} \).
The wall temperature, referred to as $T_{\text{bot}}$, is fixed at \( T_{\text{bot}} = 97.5 \, \mathrm{K} \), which corresponds to a ratio of the wall-to-recovery temperature of \( T_{\text{bot}} / T_r \simeq 0.25\). Here, the recovery temperature \( T_r \) for the flat plate is defined as:

\begin{equation}
    T_r \equiv T_\text{ref} \left[ 1 + Pr^{1/3}(\gamma - 1) M_\text{ref}^2 / 2 \right],
\end{equation}
where $Pr=0.71$, which is consistent with the definition in the previous work of \citet{zhang2018}. 

The configuration of the flat-plate DNS follows that of \citet{zhang2018}, with the key difference that an extra box is added to extend the streamwise domain from $91.7\delta_i$ to $240.8\delta_i$. This yields an increase in the friction Reynolds number $Re_\tau$ that is at least comparable to that of the Couette flow. In addition, the spanwise domain size is enlarged from $8.8\delta_i$ to $26.6\delta_i$ to accommodate the downstream growth of the boundary layer. Here, $\delta_i = 1.33$ mm denotes the boundary layer thickness at the inflow plane of Box 1, defined based on $\overline{u} = 0.99U_{\text{ref}}$. 

The turbulent statistics for the flat-plate DNS are obtained at a streamwise location $x_a = 245.1\delta_i$ downstream from the inlet of Box 1. This position is selected so that the friction Reynolds number matches that of the Couette flow, yielding $Re_{\tau} \simeq 713$. At this sampling station, the local boundary layer thickness has grown to $\delta = 6.3$ mm, as reported in table~\ref{tbl:conditions}. Additional details of the DNS can be found in \citet{zhang2018}.

\section{Numerical Method}\label{sec:Numerical method}
In both the Couette and flat-plate DNS three-dimensional fully compressible Navier--Stokes equations are solved. The viscosity (\(\mu\)) is computed using Sutherland's law, and the thermal conductivity (\(\kappa\)) is evaluated from \(\kappa = \mu C_p / Pr\). 

These two DNS calculations adopt different numerical approaches. The Couette flow is simulated using CFDSU, a sixth-order compact finite difference solver developed by \citet{nagarajan2003robust} and under continued development at Purdue. The code employs a staggered grid in which thermodynamic variables (density, pressure, temperature) are stored at cell centers, while velocity components and their associated momenta are stored at cell faces. Near the domain boundaries, the spatial accuracy is reduced to fourth order to accommodate boundary conditions. Since the Couette flow configuration does not involve shock formation, no shock-capturing method is employed. Time advancement is performed using a four-stage, third-order Strong-Stability preserving (SSP) Runge–Kutta scheme (\cite{gottlieb2001}), and a sixth-order compact filter \citep{lele1992compact} with filter coefficient $\alpha = 0.495$ is applied to maintain numerical stability.

On the other hand, the flat-plate DNS employs a high-order finite difference solver, where the inviscid fluxes are evaluated using a seventh-order weighted essentially non-oscillatory (WENO) scheme that incorporates limiters to capture eddy shocklets while simultaneously reducing numerical dissipation~\citep{jiang1996efficient}. The viscous fluxes are computed using a fourth-order central difference scheme. Time advancement is performed with a third-order low-storage Runge–Kutta method. At the top boundary, a non-reflecting, unsteady boundary condition is enforced, and periodic boundary conditions are applied in the spanwise direction. Further information on the flat-plate simulation approach can be found in \citet{zhang2017effect, zhang2018}. 

\subsection{Formalism for statistical averaging and spatial filtering operators}

Statistical averaging and spatial filtering operations are applied to the present hypersonic Couette flow and flat-plate DNS data to investigate flow statistics and sub-filter-scale (SFS) phenomena. The operators \( \langle \cdot \rangle \) and \( \{ \cdot \} \) denote Reynolds averaging and Favre (density-weighted) averaging, respectively, over time and the homogeneous directions (\(x, z\)):

\begin{equation}
\langle \phi \rangle = \frac{1}{N_t N_x N_z} \sum_{l=1}^{N_t} \sum_{k=1}^{N_z} \sum_{i=1}^{N_x} \phi_{ikl}, \qquad \{ \phi \} = \frac{\langle \rho \phi \rangle}{\langle \rho \rangle}
\label{averaging_couette_eq}
\end{equation}
where \( N_t \), \( N_x \), and \( N_z \) are the number of temporal snapshots and the number of grid points in the streamwise and spanwise directions, respectively. The values of \(N_x\) and \(N_z\) are listed in Table~\ref{tbl:conditions}. The number of snapshots used for statistical averaging is \(N_t = 200\) for the Couette flow and \(N_t = 285\) for the flat plate. In both cases, the averaging is performed over a physical time interval of \(6.5\times10^{-4}\,\mathrm{s}\).The flat-plate boundary layer is inhomogeneous in the streamwise direction, and the analysis therefore focuses on flow statistics at a specific streamwise location \(x_a\). Averaging in the streamwise direction is performed over a finite interval defined as \(x_a - 0.5\,\delta \leq x \leq x_a + 0.5\,\delta\), where \(\delta\) denotes the boundary layer thickness at \(x = x_a\).

The notations \( (\cdot)\text{'} \) and \( (\cdot)\text{"} \) denote fluctuations about the Reynolds and Favre averages, respectively:

\begin{equation}
\phi\text{'} = \phi - \langle \phi \rangle, \qquad \phi\text{"} = \phi - \{ \phi \}
\end{equation}
Spatial filtering and Favre spatial filtering are denoted by the overline \( \overline{(\cdot)} \) and the tilde \( \widetilde{(\cdot)} \), respectively:

\begin{equation}
\overline{\phi}(x) = \int \phi(\xi) G(x,\xi) \, d\xi, \qquad \widetilde{\phi} = \frac{\overline{\rho \phi}}{\overline{\rho}}
\label{filtered_phi}
\end{equation}
where \( G(x,\xi) \) is the filter kernel. For the sharp spectral filter used in SFS analysis, the kernel is defined as

\begin{equation}
G(x,\xi) = \frac{1}{\overline{\Delta}} \frac{\sin \left( \pi (\xi - x)/\overline{\Delta} \right)}{\pi (\xi - x)/\overline{\Delta}}
\label{kernel_spectral}
\end{equation}
Applying the Fourier transform yields the corresponding transfer function:
\begin{equation}
\hat{G}(k) = 
\begin{cases}
1 & \text{if } |k| \leq \pi/\overline{\Delta} \\
0 & \text{otherwise}
\end{cases}
\end{equation}
Here, \( \overline{\Delta} \) denotes the cut-off length scale of the filter. Two filter widths are considered, corresponding to four and eight times the DNS grid spacing of the Couette flow, denoted by \( \overline{\Delta}^{(1)} \) and \( \overline{\Delta}^{(2)} \). The filter widths in the streamwise and spanwise directions are denoted by \( \overline{\Delta}_x \) and \( \overline{\Delta}_z \). 
To enable a consistent comparison between the two flows, the filter widths are normalised using a characteristic length scale \( \delta_c \), defined as \( 2\delta \) for the Couette flow and \( \delta \) for the flat-plate boundary layer, with the top wall in the Couette flow serving as a surrogate for the freestream in the flat-plate boundary layer. This normalisation ensures that, although the filter widths correspond to different numbers of grid points in each flow, they represent comparable filtering scales relative to the characteristic size of the flow. The filter widths used in the present study are summarised in table~\ref{Table Filter}.

A second filter used is the top-hat filter, for which the kernel is defined as:

\begin{equation}
G(x,\xi) = 
\begin{cases}
1/\overline{\Delta} & \text{if } |x - \xi| \leq \overline{\Delta}/2 \\
0 & \text{otherwise}
\end{cases}
\label{kernel_top_hat}
\end{equation}
The transfer function for this filter is given by:
\begin{equation}
\hat{G}(k) = \frac{\sin(k \overline{\Delta}/2)}{k \overline{\Delta}/2}
\end{equation}
Using the trapezoidal rule, the top-hat filter can be discretized as:
\begin{equation}
\overline{\phi}(x_i) = \alpha \phi(x_{i - n/2}) + \beta \sum_{j = i - n/2 + 1}^{i + n/2 - 1} \phi(x_j) + \alpha \phi(x_{i + n/2})
\end{equation}
Here, $n$ denotes the ratio of the filter width to the grid spacing, i.e. $\overline{\Delta} = n \Delta$, and the weights \( \alpha \) and \( \beta \) depend on the filter width relative to the grid spacing. Using the filter widths listed in table~\ref{Table Filter}, the corresponding weights in the streamwise and spanwise directions can be determined for both the Couette and flat plate cases. Consequently, the same normalized filter width yields different values of \( \alpha \) and \( \beta \) for the two flow configurations. The specific weights employed in the present study are listed in table~\ref{tab:filter_weights} along with values of $n_x$ and $n_z$. Lastly, the notations \( (\cdot)^{\prime} \) and \( (\cdot)^{\prime \prime} \) denote residuals with respect to spatial filtering and Favre spatial filtering, respectively:

\begin{equation}
\phi^{\prime} = \phi - \overline{\phi}, \qquad \phi^{\prime \prime} = \phi - \widetilde{\phi}
\end{equation}

\begin{table}
\centering
\begin{tabular}{l c c}
Filter Width ($\overline{\Delta}$)&  $\overline{\Delta}_x/\delta_c$ & $\overline{\Delta}_z/\delta_c$ \\
$\overline{\Delta}^{(1)}$ & 0.09375 & 0.03125\\ 
$\overline{\Delta}^{(2)}$ & 0.1875 & 0.0625 \\
\end{tabular}
\caption{Filter widths applied to the hypersonic Couette flow and flat-plate boundary layer. The filter widths are normalized using the characteristic flow length scale, $\delta_c$, defined as the full channel height, \(2\delta\), for the Couette flow and the boundary-layer thickness, \(\delta\), for the flat plate.}
\label{Table Filter}
\end{table}

\begin{table}
\centering
\begin{tabular}{l c c c c c c c}
Flow & Filter Width ($\overline{\Delta}$) & $n_x$ & $n_z$ & $\alpha_x$ & $\beta_x$ & $\alpha_z$ & $\beta_z$ \\
Couette    & $\overline{\Delta}^{(1)}$ & 4  & 4  & 1/8  & 1/4  & 1/8  & 1/4  \\
Couette    & $\overline{\Delta}^{(2)}$ & 8  & 8  & 1/16 & 1/8  & 1/16 & 1/8  \\
Flat plate & $\overline{\Delta}^{(1)}$ & 11 & 6  & 1/22 & 1/11 & 1/12 & 1/6  \\
Flat plate & $\overline{\Delta}^{(2)}$ & 23 & 13 & 1/46 & 1/23 & 1/26 & 1/13 \\
\end{tabular}
\caption{Top-hat filter parameters for the Couette flow and flat-plate boundary layer.}
\label{tab:filter_weights}
\end{table}

\begin{figure}
\centering
\includegraphics[width=1\textwidth]{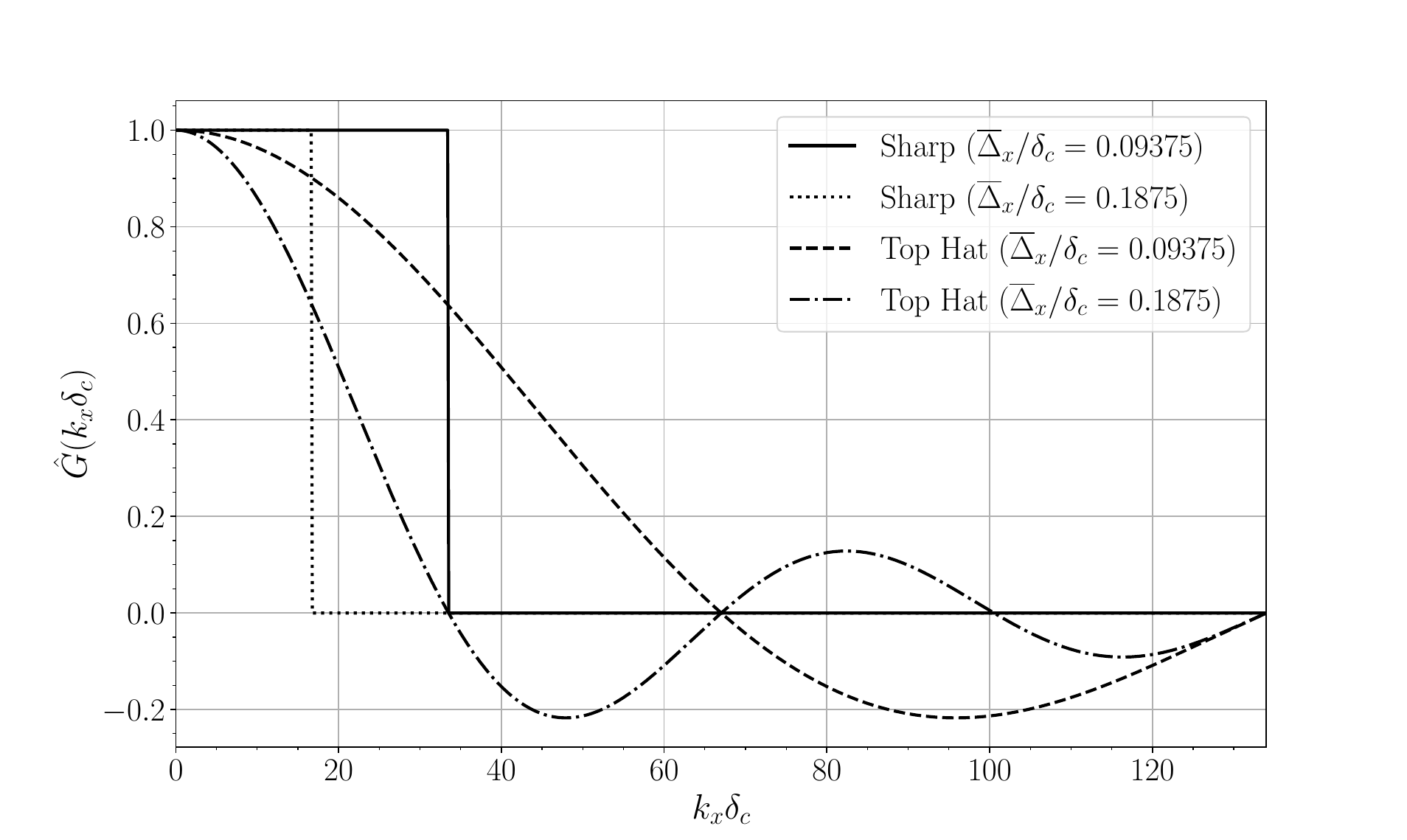}
\caption{Transfer functions of the sharp spectral and top-hat filters as functions of the streamwise wavenumber normalized by the characteristic length scale  ($k_x\delta_c$), for filter widths of $0.09375$ and $0.1875$.}
\label{Transfer function}%
\end{figure}%

To assess the sensitivity of the extracted SFS quantities to the filtering operation, two representative filters are considered: the top-hat filter and the sharp spectral filter. The top-hat filter introduces a gradual attenuation of small-scale fluctuations, leading to an effective dissipative behaviour, whereas the sharp spectral filter provides an idealised cut-off in spectral space, retaining all scales below the filter width while removing higher-wavenumber components without additional dissipation. Figure~\ref{Transfer function} presents their transfer functions for the filter widths used. In this framework, variations in the filter formulation provide an analogue of the implicit filtering arising from the numerical scheme in LES, as they modify the treatment of high-wavenumber content and the associated numerical dissipation.

\subsection{LES Formalism and SFS Analysis}
In practical LES, the effective filtering operation is generally determined implicitly by the computational mesh and numerical discretisation. In the present study, an explicit spatial filter is instead applied to the DNS data as a post-processing procedure to generate LES-equivalent filtered flow fields and the associated exact SFS terms. The filtered form of the compressible Navier–Stokes equations can be expressed as:
 
\begin{equation}
\frac{\partial \overline{\rho}}{\partial t} + \frac{\partial \overline{\rho} \widetilde{u_j}}{\partial x_j}=0, 
\end{equation}
\begin{equation}
\frac{\partial \overline{\rho} \widetilde{u_i}}{\partial t} + \frac{\partial \overline{\rho} \widetilde{u_i}\widetilde{u_j}}{\partial x_j} = -\frac{\partial \overline{p}}{\partial x_i}+\frac{\partial \overline{\sigma _{ij}}}{\partial x_j}-\frac{\partial \overline{\rho}\tau_{ij}^{\text{SFS}}}{\partial x_j},
\end{equation}
\begin{equation}
    \frac{\partial \overline{E}}{\partial t}+\frac{\partial (\overline{E}+\overline{p} ) \widetilde{u_j}}{\partial x_j}=\frac{\partial}{\partial x_j}\left( k \frac{\partial \widetilde{T}}{\partial x_j}\right)+\frac{\partial \overline{\sigma _{ij}} \widetilde{u_i}}{\partial x_j} -\frac{\partial \overline{\rho} C_p q_j^\text{SFS}}{\partial x_j} -\frac{\partial}{\partial x_j}\left(\frac{1}{2}\overline{\rho}\vartheta _j^\text{SFS} \right) +\epsilon^\text{SFS}
\end{equation}
here $\overline{E}$ is the spatial filtered total energy per unit volume and given as
\begin{equation}
    \overline{E} = \frac{\overline{p}}{\gamma -1}+\frac{1}{2}\overline{\rho}\widetilde{u_k}\widetilde{u_k}+\frac{1}{2}\overline{\rho}\tau_{kk}^{\text{SFS}}.
\end{equation}
and $\overline{\sigma_{ij}}$ is the filtered viscous stress tensor defined as 
\begin{equation}
\overline{\sigma_{ij}}=\overline{\mu\left( \frac{\partial u_i}{\partial x_j} +\frac{\partial u_j}{\partial x_i} - \frac{2}{3} \frac{\partial u_k}{\partial x_k} \delta_{ij} \right)}
\label{eq: filtered viscous stress tensor}
\end{equation}

The $\tau _{ij}^\text{SFS}$, $q_j^\text{SFS}$, $\vartheta_j^\text{SFS}$ and $\epsilon^\text{SFS}$ are the unclosed sub-filter scale (SFS) stress tensor, the SFS temperature flux, the SFS kinetic energy advection and the SFS turbulent heat dissipation. The expressions for the SFS stress tensor and heat flux are mentioned below.
\begin{equation}
    \tau_{ij}^\text{SFS} = \widetilde{u_i u_j} -\widetilde{u_i}\widetilde{u_j},
    \label{tau_def}
\end{equation}
\begin{equation}
    q_j^\text{SFS} = \widetilde{Tu_j}-\widetilde{T}\widetilde{u_j}.
\end{equation}
In the present paper, the sub-filter scale (SFS) shear stress $\tau_{12}^{\mathrm{SFS}}$, the trace of the SFS stress tensor $\tau_{\mathrm{Trace}}^{\mathrm{SFS}}$, and the wall-normal SFS heat flux $q_2^{\mathrm{SFS}}$ are analyzed for both the hypersonic Couette flow and the flat-plate boundary layer. The expressions for $\vartheta_j^{\mathrm{SFS}}$ and $\epsilon^{\mathrm{SFS}}$ are given by

\begin{equation}
    \vartheta_j^\text{SFS} = \widetilde{u_k u_k u_j} -\widetilde{u_k}\widetilde{u_k}\widetilde{u_j},
\end{equation}
\begin{equation}
    \epsilon^\text{SFS} = \frac{\partial \overline{\sigma _{ij}u_i}}{\partial x_j}-\frac{\partial \overline{\sigma _{ij}} \widetilde{u_i}}{\partial x_j}.
    \label{theta_def}
\end{equation}

In the present analysis, explicit spatial filtering is applied to the DNS data in order to extract the SFS contributions. This filtering operation is defined independently of any numerical discretisation and is used solely as a post-processing procedure.

\begin{figure}
\centering
\includegraphics[width=1\textwidth]{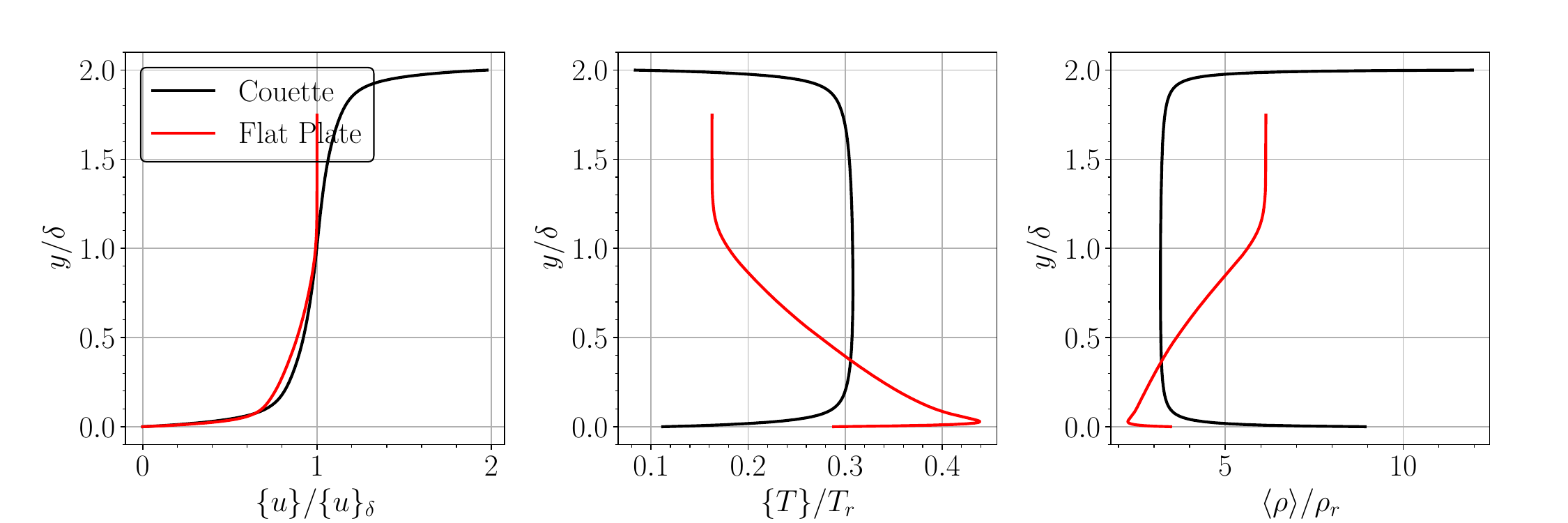}
\put(-370,120){(a)}
\put(-250,120){(b)}
\put(-130,120){(c)}
\caption{Normalized mean profiles of (a) streamwise velocity, (b) temperature and (c) density vs $y/\delta$ for the hypersonic Couette flow and flat plate. $\delta$ here is the half-channel width for Couette and the boundary layer thickness at $x_a$ for the flat plate.}
\label{Mean y/delta avg}%
\end{figure}%

\section{Time-averaged DNS statistics}\label{sec:time-average}

Time-averaged statistics of the DNS for Couette and flat-plate configurations are examined in this section. Mean flow profiles, temperature--velocity relations, and transformed velocity and temperature profiles are analysed in \S~\ref{Mean flow profiles}. Second-order statistics, including Reynolds shear stress and temperature, density, and velocity fluctuations, are discussed in \S~\ref{second-order statistics}. In \S~\ref{sec:spectra}, the streamwise and spanwise energy spectra for both the Couette and flat plate flows are presented and analysed. This analysis of the time-averaged DNS statistics highlights the similarities and differences in the mean flow and turbulence characteristics of the two configurations, thereby providing the foundation for the subsequent SFS analysis presented in \S~\ref{SFS analysis}.

\subsection{Mean Flow Profiles \label{Mean flow profiles}}
Normalized mean flow profiles of streamwise velocity, temperature, and density are presented in figure~\ref{Mean y/delta avg} as functions of the wall-normal coordinate $y/\delta$. The velocity, temperature, and density profiles are normalized by the boundary-layer edge velocity, recovery temperature $T_r$, and corresponding density $\rho_r$, respectively. The density, $\rho_r$, corresponding to the recovery temperature, $T_r$, defined in equation~(\ref{eq:recovery}), is computed as
$\rho_r=\rho_\mathrm{ref}T_{\mathrm{ref}}/{T_r}$. For both the flat plate and Couette flow configurations, the streamwise velocity exhibits a monotonic increase, and the profiles remain similar in shape till $y/\delta=1$. Strong near-wall gradients are observed in both cases due to the dominance of viscous shear stress. At $y/\delta = 1$, the gradient $\partial u / \partial y \approx 0$ for the flat plate, as the velocity reaches the freestream value at the boundary layer edge. In contrast, the Couette velocity profile remains antisymmetric, and its slope at $y/\delta = 1$ is non-zero owing to the wall-driven shear.

Despite the similarity in the velocity profiles, the density and temperature fields differ significantly between the flat-plate boundary layer and the Couette configuration. In the Couette case, the temperature increases monotonically from the bottom wall to \(y/\delta \approx 0.2\), remains nearly constant over the channel core, and then decreases toward the upper wall, reaching the prescribed upper-wall temperature. This behaviour is consistent with observations in supersonic and hypersonic channel flows~\citep{ChenS_PRF_2021,ChenS_JFM_2021}, where both walls are maintained below the recovery temperature. In contrast, the flat-plate temperature profile increases sharply near the wall, reaching a maximum at \(y/\delta \approx 0.02\), and subsequently decreases towards the freestream value at the boundary-layer edge. This difference reflects the distinct mechanisms governing thermal transport in the two flows. In Couette flow, viscous dissipation occurs near both walls, and the resulting heat is retained within the channel, leading to an accumulation of thermal energy and a relatively uniform temperature in the channel core. In the flat-plate boundary layer, however, viscous dissipation is confined to the region near the bottom wall, resulting in a local temperature maximum, above which the temperature decreases as the influence of viscous heating diminishes towards the freestream. The density profiles follow the temperature distributions in both flows given that the pressure is nearly constant across the channel or boundary layer. In the Couette case, the density decreases from the bottom wall to \(y/\delta \approx 0.2\), remains nearly constant up to the channel centerline, and then increases toward the upper wall, whereas in the flat-plate case a minimum is observed at approximately the same wall-normal location as the temperature maximum, followed by a gradual recovery towards the freestream value.

\begin{table}
\centering
\begin{tabular}{l c}
Correlation & $r_g$\\ \\
Crocco-Busemann & 1\\ 
Walz &r\\
Huang \textit{et al.} &$Pr_t$\\
Duan and Martin &$r[a + (1-a)(\{T_w\}-\{T_e\})/(T_r-\{T_e\})]$\\
Zhang \textit{et al.} &$2C_p(\{T_w\}-\{T_e\})/\{u_e\}^2-2Prq_w/(\{u_e\}\{\tau_w\})$\\
\end{tabular}
\caption{A list of various temperature-velocity relations. $r\equiv(T_r-T_e)/(T_0-T_e)$ here is the recovery factor and the turbulent Prandtl number $Pr_t$ is 0.9. The parameter $a = 0.8259$ for the model by \cite{duan2011b}.}
\label{Temperature Velocity Correlation}
\end{table}
\begin{figure}
\centering
\includegraphics[width=1\textwidth]{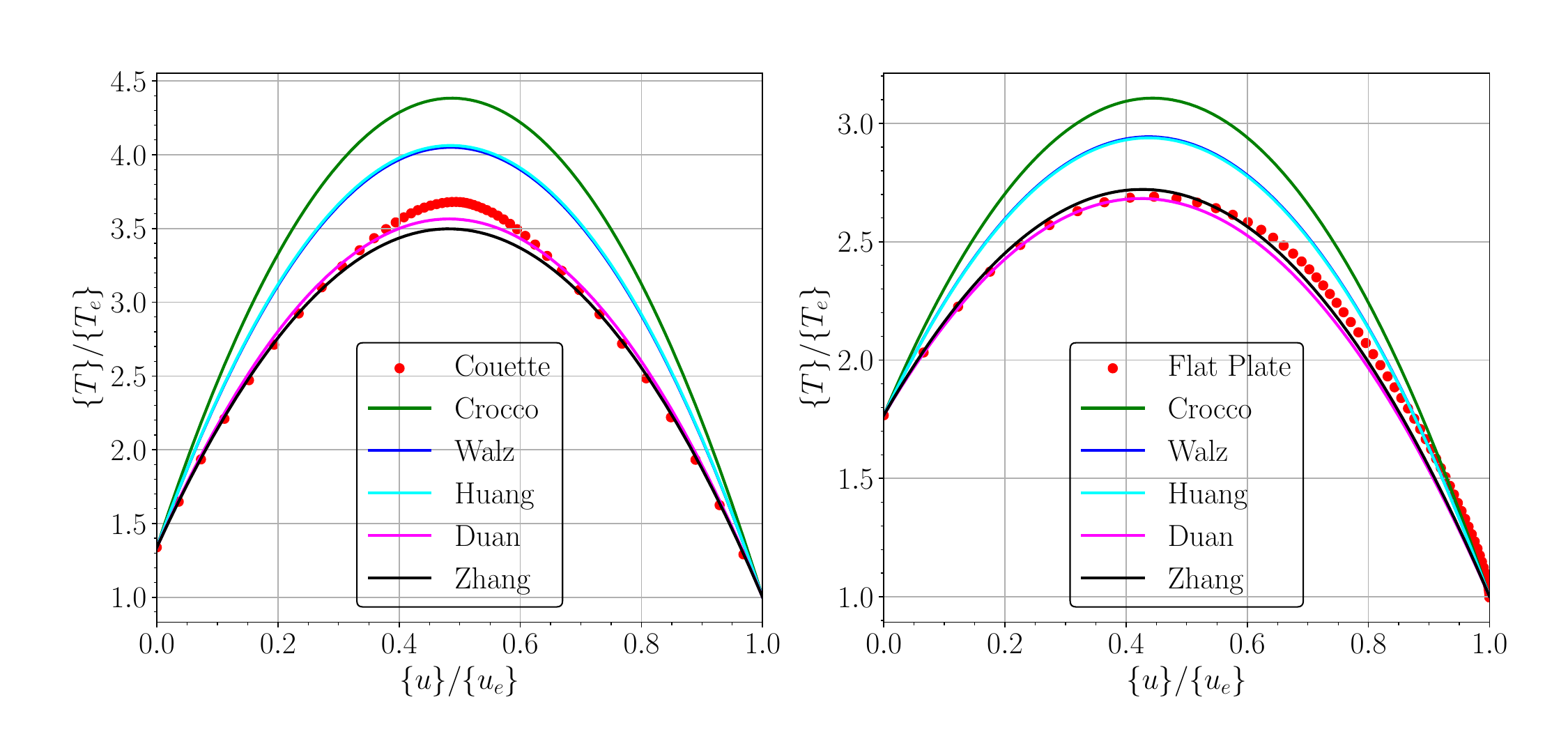}
\put(-370,160){(a)}
\put(-185,160){(b)}
\caption{Comparison of temperature-velocity relation for (a) Couette and (b) Flat Plate at $x_a/\delta_i = 69.73$ using several correlations.}
\label{Temperature VelocityCorr}%
\end{figure}%

The comparison of the mean profiles reveals a discrepancy in the temperature distributions between the flat-plate boundary layer and the Couette flow. While existing temperature–velocity correlations have been shown to remain valid for hypersonic boundary layers~\citep{zhang2018, huang2022, toki_JFM_2025}, their applicability to Couette flow remains unclear. Several correlations are therefore applied to both configurations to assess their validity, particularly for the Couette flow.

The temperature–velocity relation is generally written as
\begin{equation}
    \{T\} = \{T_w\}-\left(\{T_w\}-\{T_e\}-\frac{r_g}{2C_p}\{u_e\}^2\right)\frac{\{u\}}{\{u_e\}}-\frac{r_g}{2C_p}\{u\}^2
\end{equation}
Here, the parameter $r_g$ depends on the specific correlation, and the subscript $e$ denotes the value at the moving wall in the Couette flow and at the boundary-layer edge for the flat plate. The correlations considered are summarised in table~\ref{Temperature Velocity Correlation}, including those based on the Crocco–Busemann relation \citep{busemann1931handbuch,crocco1932sulla}, the Walz relation \citep{walz1959compressible}, the Huang relation \citep{huang1993skin}, the modified relation by Duan and Martin \citep{duan2011b}, and the generalized Walz relation proposed by \citet{zhang2014generalized}.

These correlations are compared in figure \ref{Temperature VelocityCorr}. For the Couette flow, the Crocco–Busemann, Walz~\citep{walz1959compressible}, and Huang~\citep{huang1993skin} correlations overestimate the temperature–velocity relation, whereas the correlations by \citet{duan2011b} and \citet{zhang2014generalized} show good agreement except in the intermediate range ($0.4 \lesssim \{u\}/\{u_e\} \lesssim 0.6$), where they underestimate the temperature–velocity relation. For the flat-plate boundary layer, the Crocco–Busemann, Walz, and Huang correlations similarly overestimate the temperature–velocity relation. The correlations by \citet{duan2011b} and \citet{zhang2014generalized} again agree well with the DNS results, except for higher values of $\{u\}/\{u_e\}$ ($\{u\}/\{u_e\} \gtrsim 0.5$), where they tend to underestimate the temperature–velocity relation. Overall, the modified correlations by \citet{duan2011b} and \citet{zhang2014generalized} show the best agreement with the DNS data for both configurations, suggesting that such modifications are essential for accurately capturing the temperature–velocity relation under hypersonic conditions. 

The deviation from the DNS data is, however, more pronounced in the Couette flow than in the flat-plate boundary layer, particularly in the intermediate range ($0.4 \lesssim \{u\}/\{u_e\} \lesssim 0.6$). This range corresponds to the core region, where the temperature is nearly constant. Since such a constant-temperature behaviour is not observed in the flat-plate boundary layer, it can be inferred that correlations calibrated for boundary-layer flows are less effective for the Couette configuration. Nevertheless, the correlations by \citet{duan2011b} and \citet{zhang2014generalized} show good agreement in both the low-velocity ($\{u\}/\{u_e\} \lesssim 0.2$) and high-velocity ($\{u\}/\{u_e\} \gtrsim 0.8$) ranges. These ranges correspond to the vicinity of the bottom and top walls, respectively, indicating that these correlations remain applicable in the near-wall regions even for internal flows such as Couette flow.

To further examine the differences in the mean temperature and velocity profiles, scaling laws are employed. To account for wall-normal property variations, several velocity scaling laws \citep{Trettel2016, volpiani_PRF_2020, griffin_PNAC_2021, bai2022, hasan_PRF_2023} and temperature scaling laws \citep{patel_PRF_2017, toki_JPP_2020, wan_pof_2020, huang2023velocity} have been proposed in the literature. In the present study, the transformations by \citet{Trettel2016}, \citet{huang2023velocity}, \citet{griffin_JFM_2023} and \cite{hasan_PRF_2023} are adopted for the velocity and temperature profiles, respectively.

The transformation of \citet{Trettel2016} is given by
\begin{equation}
    u^+_\text{TL} = \int ^{u^+} _0 \left( \frac{\langle \rho \rangle}{\langle \rho _\text{bot} \rangle} \right) ^{1/2} \left[ 1 + \frac{1}{2} \frac{1}{\langle \rho \rangle} \frac{\partial \langle \rho \rangle}{\partial y} y -\frac{1}{\langle \mu \rangle} \frac{\partial \langle \mu \rangle}{\partial y} y \right] du^+
\end{equation}
where $u^{+}\equiv \langle u \rangle/u_\tau$. The transformation of \citet{huang2023velocity} is given by
\begin{equation}
    T^+_\text{TL} = \int ^{\theta} _0 \frac{1}{B_q + (\gamma - 1)M^2_{\tau}u^+}\left( \frac{\langle \rho \rangle}{\langle \rho _\text{bot} \rangle} \right) ^{1/2} \left[ 1 + \frac{1}{2} \frac{1}{\langle \rho \rangle} \frac{\partial \langle \rho \rangle}{\partial y} y -\frac{1}{\langle \mu \rangle} \frac{\partial \langle \mu \rangle}{\partial y} y \right] d\theta
\end{equation}
Here, $B_q = \langle q_\text{bot}\rangle/(\langle \rho_\text{bot}\rangle u_{\tau} c_p  T_{\text{bot}})$, $M_\tau = u_{\tau}/\left(\gamma R T_\text{bot} \right)^{1/2}$, and $\theta = (T_\text{bot} - \{T\})/T_\text{bot}$. Both transformations are derived following a similar framework and share a common structure in their integrands, allowing a consistent comparison of the transformed profiles. In the present study, the performance of these transformations is assessed for both external and internal compressible turbulent flows at the same friction Reynolds number, \(Re_\tau\).

\citet{griffin_JFM_2023} further proposed a velocity transformation for compressible wall-bounded turbulent flows that accounts for the distinct effects of compressibility on the viscous and turbulent shear stresses. This approach, referred to as the total-stress-based transformation, is expressed as
\begin{equation}
    u^+_{\text{TS}} = \int_0^{u^+} \frac{S^+_{\text{eq}}}{1+S^+_{\text{eq}}-S^+_{\text{TL}}}dy^*
\end{equation}
where $S^+_{\mathrm{eq}}=(\mu_{\mathrm{bot}}/\langle\mu\rangle)(\partial u^+/\partial y^*)$, $S^+_{\mathrm{TL}}=(\langle\mu\rangle/\mu_{\mathrm{bot}})(\partial u^+/\partial y^+)$, $y^*=y/\delta_v^*$, and $y^+=y/\delta_v$. Here, $\delta_v^*$ is the semi-local viscous length scale \citep{huang1995compressible}, defined as $\delta_v^*=\langle\mu\rangle/(\langle\rho\rangle u_\tau^*)$, where $u_\tau^*=\sqrt{\tau_{\mathrm{bot}}/\langle\rho\rangle}$, and $\delta_v$ is the viscous length scale, defined as $\delta_v=\mu_{\mathrm{bot}}/(\langle\rho_{\mathrm{bot}}\rangle u_\tau)$.

More recently, \citet{hasan_PRF_2023} proposed a velocity transformation that accounts for variable-property and intrinsic compressibility effects. The transformation is expressed as

\begin{equation}
    u^+_\text{H} = \int_0^{u^+}\left(\frac{1+\kappa y^*D^c}{1+\kappa y^*D^i}\right)\left(1-\frac{y}{\delta_v^*}\frac{d\delta_v^*}{dy}\right)\sqrt{\frac{\langle \rho \rangle}{\langle\rho_\text{bot}\rangle}} du^+
\end{equation}
where $\kappa=0.41$, $D^c=(1-\text{exp}\left(-y^*/(A^++f(M_\tau))\right))^2$ and $D^i=(1-\text{exp}\left(-y^*/A^+\right))^2$. $A^+=17$ and $f(M_{\tau}) = 19.3 M_{\tau}$.

\begin{figure}
\centering
\includegraphics[width=1\textwidth]{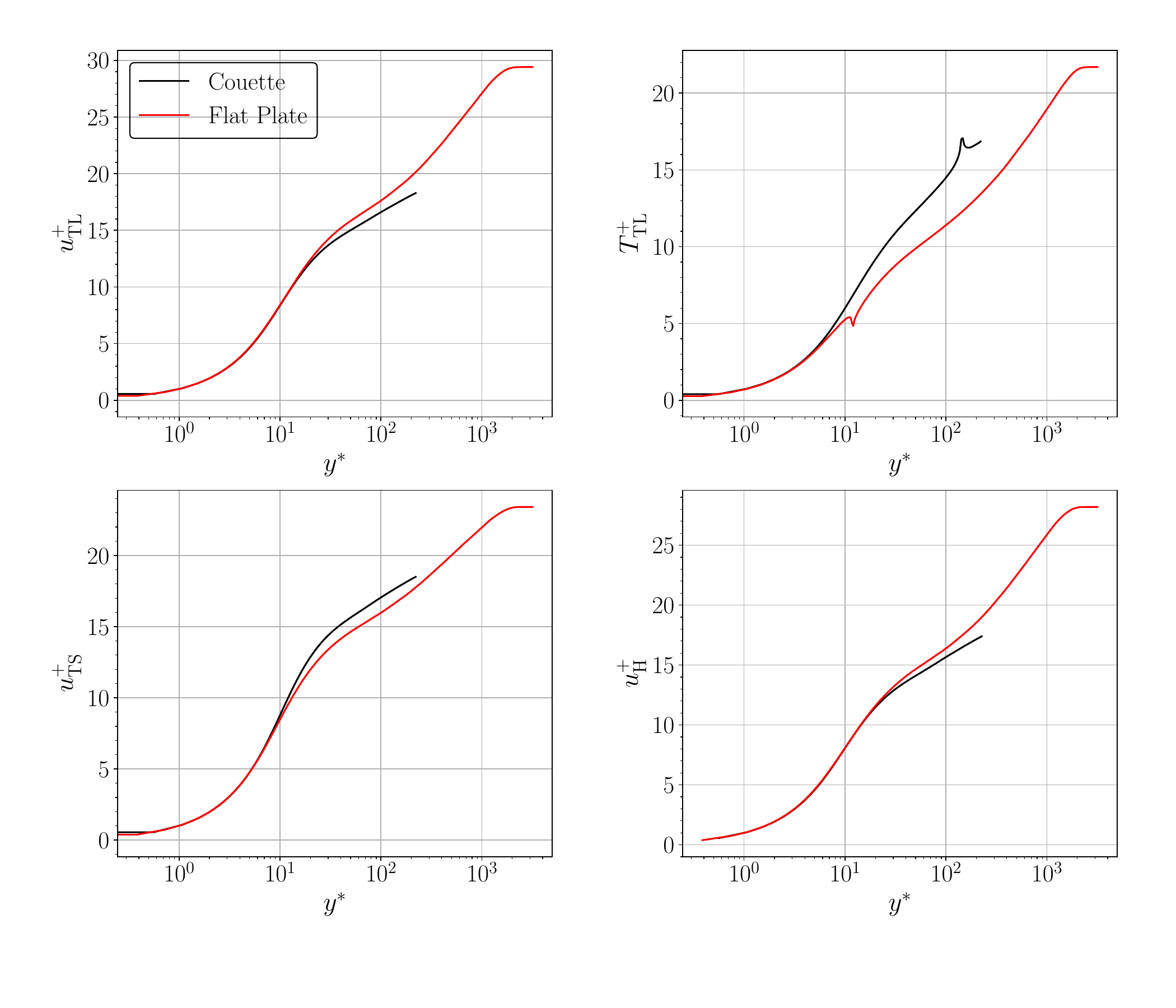}
\put(-370,315){(a)}
\put(-180,315){(b)}
\put(-370,170){(c)}
\put(-180,170){(d)}
\caption{Transformed profiles of (a) velocity via \cite{Trettel2016}'s transformation, (b) temperature via \cite{huang2023velocity}'s transformation, (c) velocity via  \cite{griffin_JFM_2023}'s and (d) velocity via \cite{hasan_PRF_2023}'s transformation.}

\label{TL transformation}%
\end{figure}

Figure \ref{TL transformation} compares the transformed profiles between the Couette and flat-plate configurations. The profiles for the Couette flow are shown only from the bottom wall to the channel centreline. The Couette and flat-plate profiles collapse onto each other for both velocity and temperature for $y^* \lesssim 10$, indicating that the same scaling laws are applicable in the viscous sublayer for both configurations. However, the velocity profile begins to deviate around $y^* = 20$, while the temperature profile deviates for $y^* \gtrsim 10$, suggesting that these transformations cannot fully account for differences between internal and external flows and are only effective up to the buffer layer. 

The temperature profile $T^+_\text{TL}$ exhibits cusps due to the change in sign of the temperature gradient. For the flat plate, the cusp occurs at $y^* \approx 12$, whereas for the Couette flow it appears at $y^* \approx 145$. The pronounced shift in the cusp location between the two flows highlights fundamental differences in their thermal structure, which are not fully collapsed by the transformation.

\subsection{Second-Order Statistics}\label{second-order statistics}

\begin{figure}
\centering
\includegraphics[width=1\textwidth]{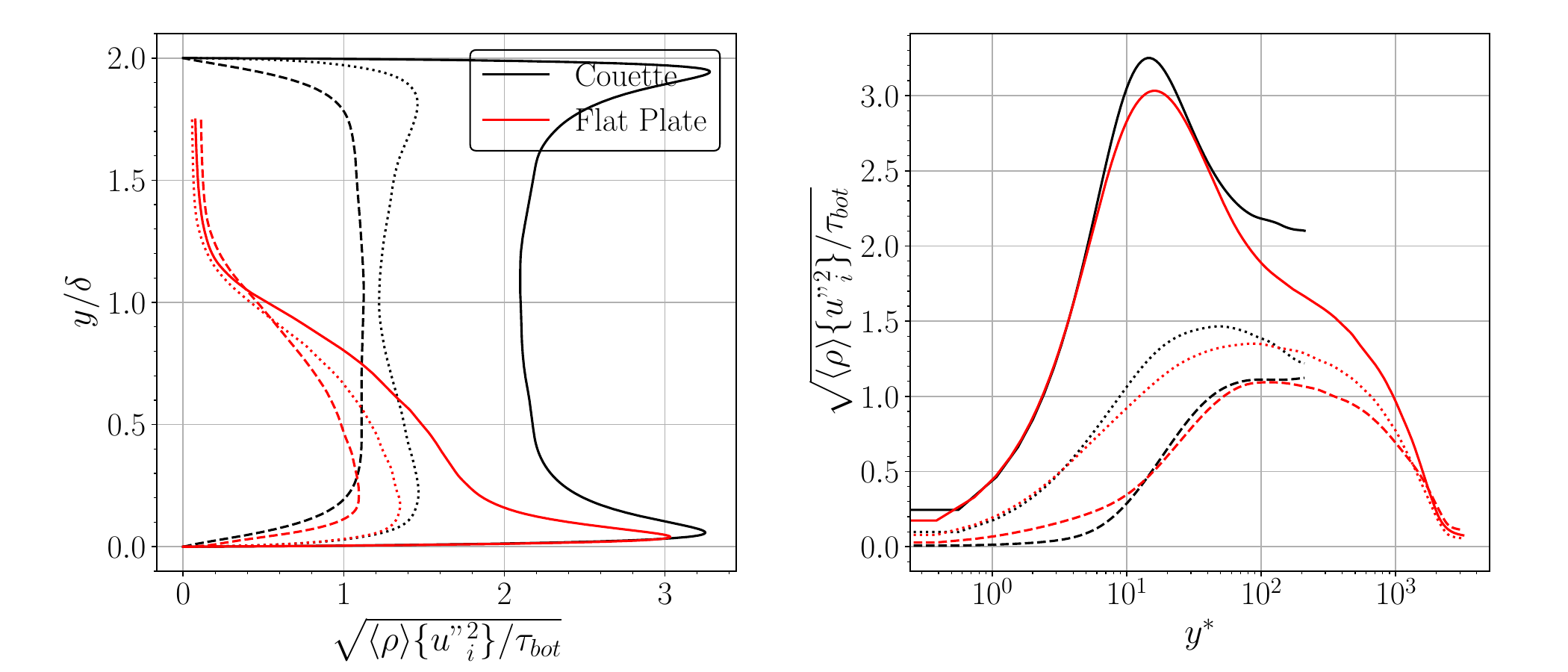}
\put(-370,160){(a)}
\put(-180,160){(b)}
\caption{RMS Velocity fluctuations as a function of (a) $y/\delta$ and (b) $y^*$ for the hypersonic Couette flow (black) and flat plate (red). Solid (--) indicates $u"_\text{rms}$, dashed (- - -) indicates $v"_\text{rms}$ and dotted $(\cdots)$ indicates $w"_\text{rms}$.}
\label{Velocity Corr}%
\end{figure}

In this section, the differences in second-order statistics are examined. Figure~\ref{Velocity Corr} presents the root mean square (RMS) velocity fluctuations for the Couette flow and the flat-plate boundary layer. The profiles are non-dimensionalised using the wall shear stress, and are shown as functions of $y/\delta$ and $y^*$. For the Couette flow, the $y^*$-scaled profiles are shown only from the bottom wall to the channel centreline. For both configurations, the streamwise velocity fluctuation is the strongest, followed by the spanwise and wall-normal components, reflecting the redistribution of turbulent kinetic energy among the velocity components.

\begin{figure}
\centering
\includegraphics[width=1\textwidth]{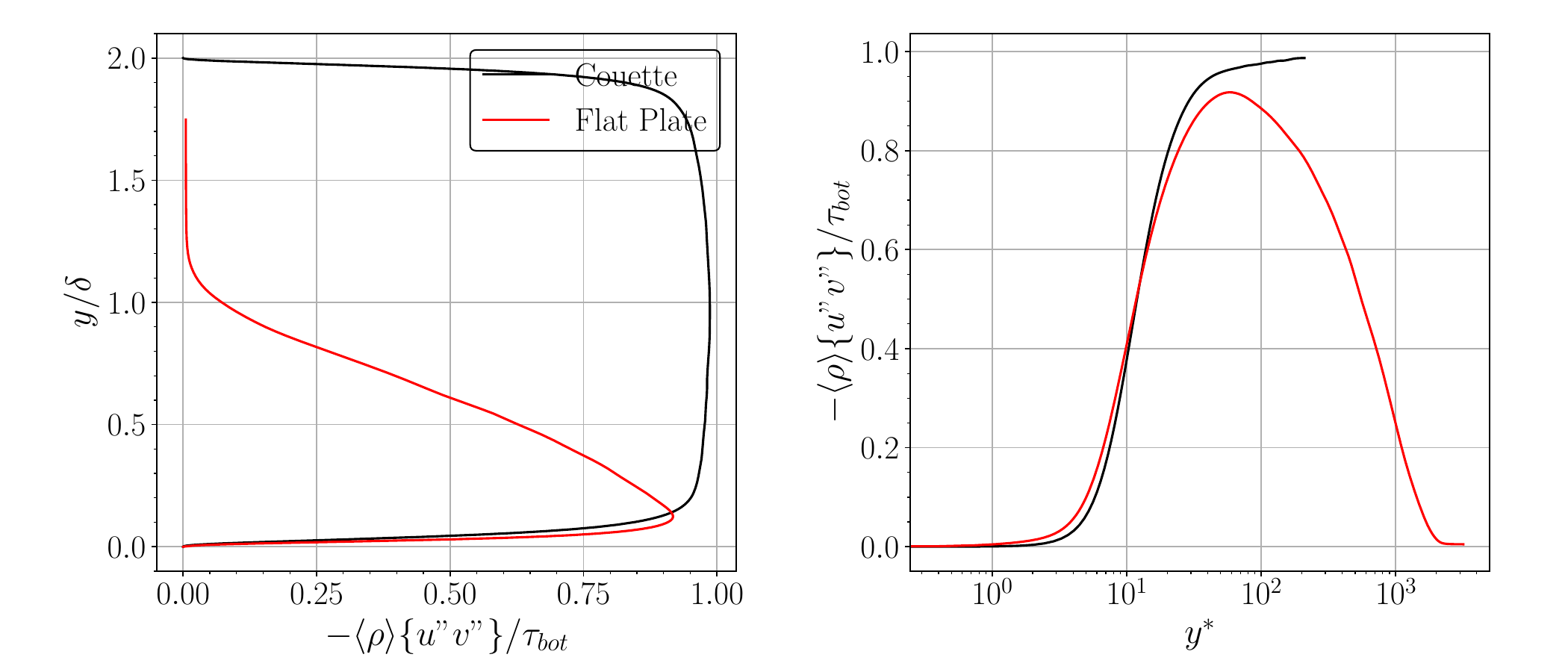}
\put(-370,160){(a)}
\put(-180,160){(b)}
\caption{Reynolds shear stress profiles as a function of (a) $y/\delta$ and (b) $y^*$ for the hypersonic Couette flow and flat plate.}
\label{Reynolds shear}%
\end{figure}%

For the Couette flow, the streamwise velocity fluctuation increases rapidly below $y/\delta = 0.05$ and reaches a peak in the buffer layer ($y^* \approx 20$), where turbulence production is strongest. It then gradually decreases to a nearly constant value over the channel core before increasing again near the upper wall and finally dropping to zero at the wall. Away from the near-wall region, turbulence production decreases, and energy is redistributed to the other velocity components. Nevertheless, the streamwise fluctuation remains approximately constant over $0.4 \lesssim y/\delta \lesssim 1.6$ owing to the non-zero mean velocity gradient across the channel core. The spanwise velocity fluctuation increases up to $y/\delta \approx 0.1$, remains nearly constant up to the channel centre, increases again towards the upper wall, and finally decreases to zero at the wall. The wall-normal component exhibits a similar trend up to $y/\delta \approx 0.2$, then remains nearly constant before decreasing to zero at the upper wall.

For the flat-plate boundary layer, the streamwise velocity fluctuation similarly increases sharply up to $y/\delta \approx 0.04$, after which it decreases monotonically towards the boundary-layer edge as it is no longer sustained by turbulence production. The spanwise and wall-normal components increase up to $y/\delta \approx 0.1$ and $y/\delta \approx 0.2$, respectively, and then gradually decrease towards the freestream.

When plotted using semi-local scaling, the velocity fluctuation profiles exhibit qualitatively similar behaviour between the Couette and flat-plate configurations, particularly in the near-wall region. However, the profiles do not fully collapse in the outer region due to the fundamental difference in outer-layer structure. Unlike the flat-plate boundary layer, the Couette flow does not possess a distinct outer layer; instead, turbulence persists throughout the domain, and the velocity fluctuations remain finite rather than approaching zero.

Figure~\ref{Reynolds shear} presents the Reynolds shear stress profiles for the Couette and flat-plate flows, shown as functions of $y/\delta$ and $y^*$. The profiles are non-dimensionalised using the wall shear stress. For the Couette flow, the Reynolds shear stress increases monotonically from the wall up to $y/\delta \approx 0.2$, after which it remains nearly constant over the range $0.2 \lesssim y/\delta \leq 1.8$ and then decreases as the upper wall is reached. This behaviour reflects the influence of the moving top wall, which maintains a fully turbulent state throughout the channel, including the core region. In contrast, for the flat-plate boundary layer, the Reynolds shear stress reaches a maximum at approximately $y/\delta \approx 0.1$ and then decreases steadily towards the boundary-layer edge. This decline is associated with the reduction in the mean velocity gradient, leading to diminished turbulence production as the outer region is approached. When plotted against $y^*$, the Reynolds shear stress profiles of the two flows show agreement in the near-wall region, whereas they deviate in the outer region, reflecting the different outer-layer structures of internal and external flows.

\begin{figure}
\centering
\includegraphics[width=1\textwidth]{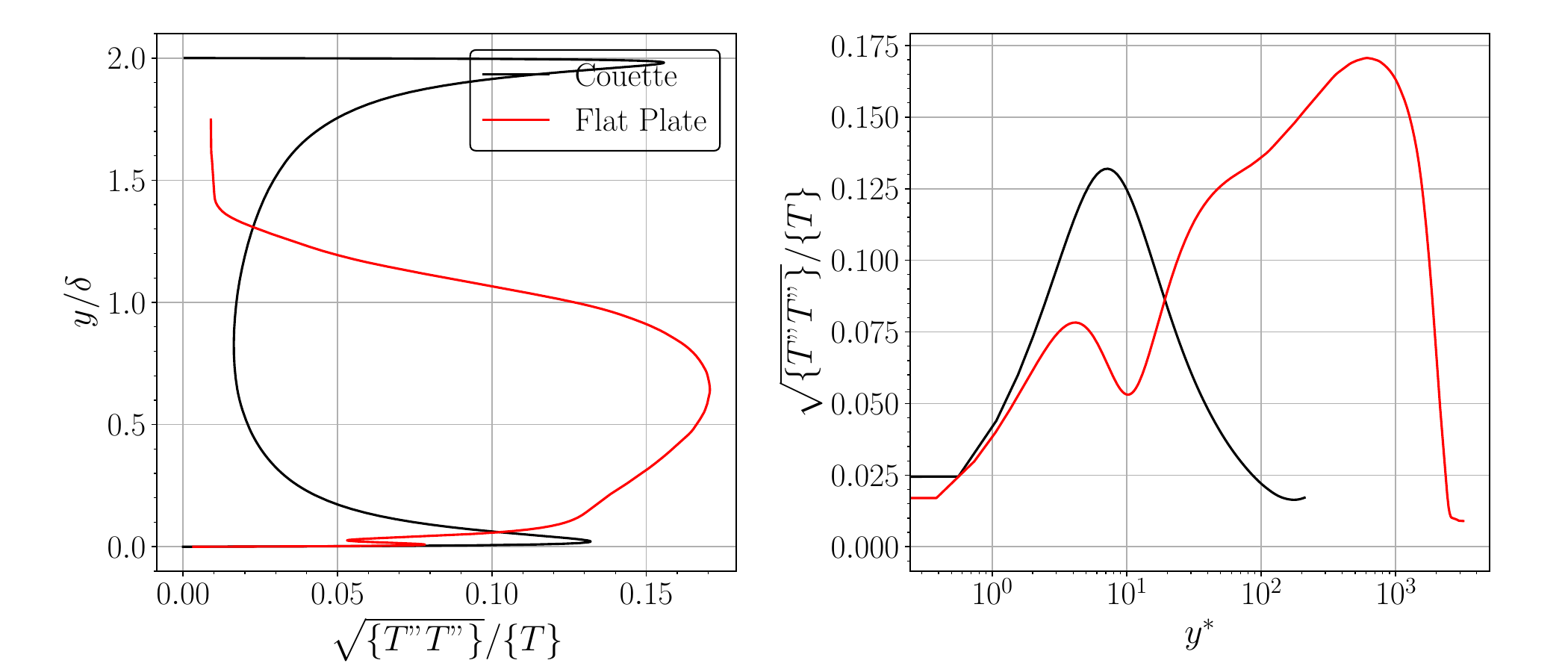}
\put(-370,160){(a)}
\put(-180,160){(b)}
\caption{RMS temperature fluctuation as a function of (a) $y/\delta$ and (b) $y^*$ for the hypersonic Couette flow and flat plate.}
\label{Temperature Corr}%
\end{figure}%

\begin{figure}
\centering
\includegraphics[width=1\textwidth]{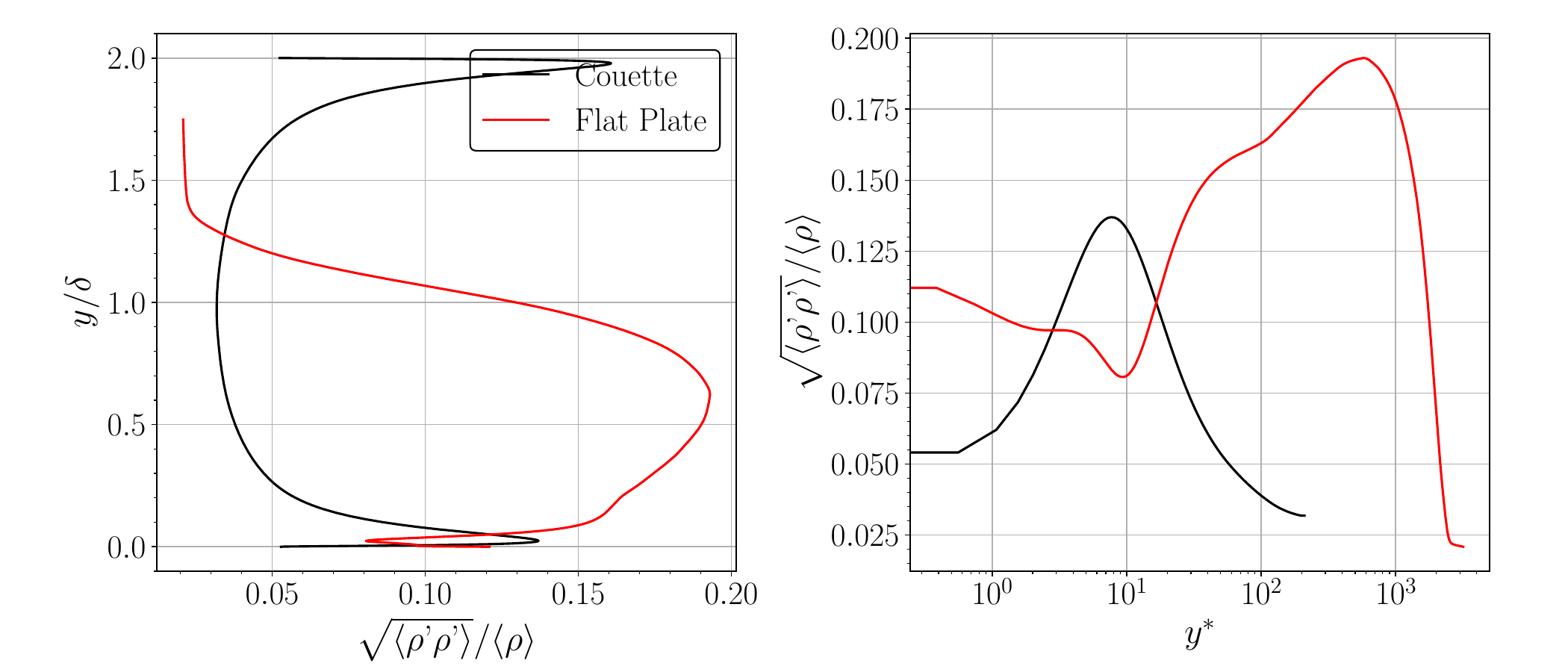}
\put(-370,160){(a)}
\put(-180,160){(b)}
\caption{RMS density fluctuation as a function of (a) $y/\delta$ and (b) $y^*$ for the hypersonic Couette flow and flat plate. }
\label{Density Corr}%
\end{figure}%

Figures~\ref{Temperature Corr} and~\ref{Density Corr} depict the RMS  temperature and density fluctuations as functions of $y/\delta$ and $y^*$. These quantities are non-dimensionalised using the local mean values at each $y$ location. For the Couette flow, the temperature fluctuation intensity increases initially, reaching a peak at $y/\delta \approx 0.02$, before decreasing toward the channel centre. It then increases again as the upper wall is approached and finally drops to zero at the wall. The near-wall peaks are associated with the strong mean temperature gradients, where wall-normal turbulent motions transport fluid across the thermal gradient and generate large temperature fluctuations. The decrease away from the walls is associated with the weakening of the mean temperature gradient, while the fluctuation intensity remains approximately constant throughout the channel core.

In contrast, the temperature and density fluctuation correlations for the flat-plate boundary layer exhibit a negative peak at $y/\delta \approx 0.05$ ($y^* \approx 10$), where the mean temperature gradient locally approaches zero, and the turbulent heat flux is reduced. As shown in figure \ref{Mean y/delta avg}, the mean temperature decreases monotonically for $y/\delta \gtrsim 0.05$. Accordingly, the temperature fluctuation remains significant throughout the boundary layer and eventually decreases to zero near $y^* \simeq 1000$ (or $y/\delta \simeq 1.6$). The density fluctuation shows a similar trend to the temperature fluctuation outside the viscous sublayer. However, within the viscous sublayer, the density fluctuation remains finite even though the temperature fluctuation diminishes towards the wall. This is due to the isothermal boundary condition imposed at the wall, which suppresses temperature fluctuations, while the pressure is not constrained to a fixed value. As a result, density fluctuations persist at the wall in response to pressure fluctuations.

From figures~\ref{Velocity Corr} to~\ref{Density Corr}, it is observed that the velocity fluctuations show qualitative agreement in the near-wall region, while discrepancies arise above the buffer layer between the Couette and flat-plate configurations. In contrast, the temperature and density fluctuations differ significantly over most of the domain, highlighting the distinct thermal behaviour in the two flows.

Since the present work aims to understand the extent to which turbulence modelling approaches developed for internal flows can be extended to external flows, it is important to examine the relationship between velocity and temperature in these two flow configurations. In particular, many compressible turbulence models are motivated by the Strong Reynolds Analogy (SRA) proposed by \citet{morkovin1962effects}. Assuming negligible total temperature fluctuations, the SRA for a zero-pressure-gradient adiabatic turbulent boundary layer can be expressed as
\begin{equation}
    Pr_t = \frac{\langle \rho\rangle\{u"v"\}\partial\{T\}/\partial y}{\langle \rho\rangle\{v"T"\}\partial\{u\}/\partial y} \approx1
\end{equation}

\begin{equation}
    -\text{R}_{u"T"} = \frac{\{u"T"\}}{\sqrt{\{u"^2\}}\sqrt{\{T"^2\}}} \approx 1
\end{equation}

\begin{equation}
    \frac{\sqrt{\{T"T"\}}/\{T\}}{(\gamma-1)M^2(\sqrt{\{u"u"\}}/\{u\})} \approx 1
    \label{eq:SRA}
\end{equation}
where $M^2=\{u\}^2/(\gamma R_\text{gas}\{T\})$.

\begin{figure}
\centering
\includegraphics[width=1\textwidth]{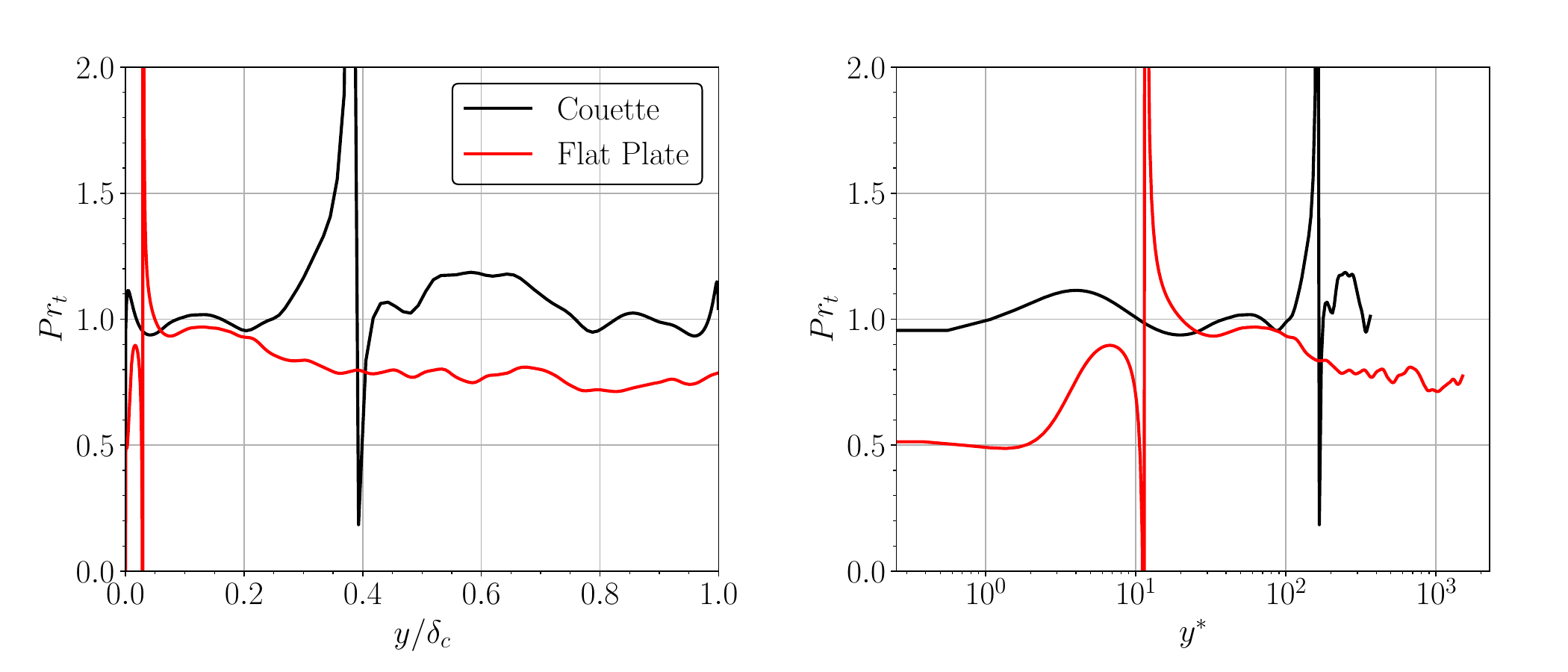}
\put(-370,157){(a)}
\put(-180,157){(b)}
\caption{Turbulent Prandtl number as a function of the (a) outer scaling $y/\delta_c$ and (b) the semi-local wall unit $y^*$. Here $\delta_c$ is $2\delta$ for the Couette and the boundary-layer thickness $\delta$, for the flat plate.}%
\label{Turbulent Prandtl}%
\end{figure}

Figure~\ref{Turbulent Prandtl} presents the turbulent Prandtl number, $Pr_t$, for the Couette and flat-plate flows as functions of $y/\delta_c$ and $y^*$. For the Couette flow, $Pr_t$ remains close to unity over most of the wall-normal extent, except in the region $0.25 \lesssim y/\delta_c \lesssim 0.35$, where it exhibits a pronounced overshoot associated with a change in the sign of the turbulent wall-normal heat flux, $\{v"T"\}$. For the flat plate, $Pr_t$ remains below unity in the near-wall region before exhibiting an overshoot in the buffer layer, again corresponding to the sign change in $\{v"T"\}$. Farther from the wall, it decreases and remains close to unity over $30 \lesssim y^* \lesssim 100$, before decreasing again through the logarithmic and outer regions of the boundary layer.

\begin{figure}
\centering
\includegraphics[width=1\textwidth]{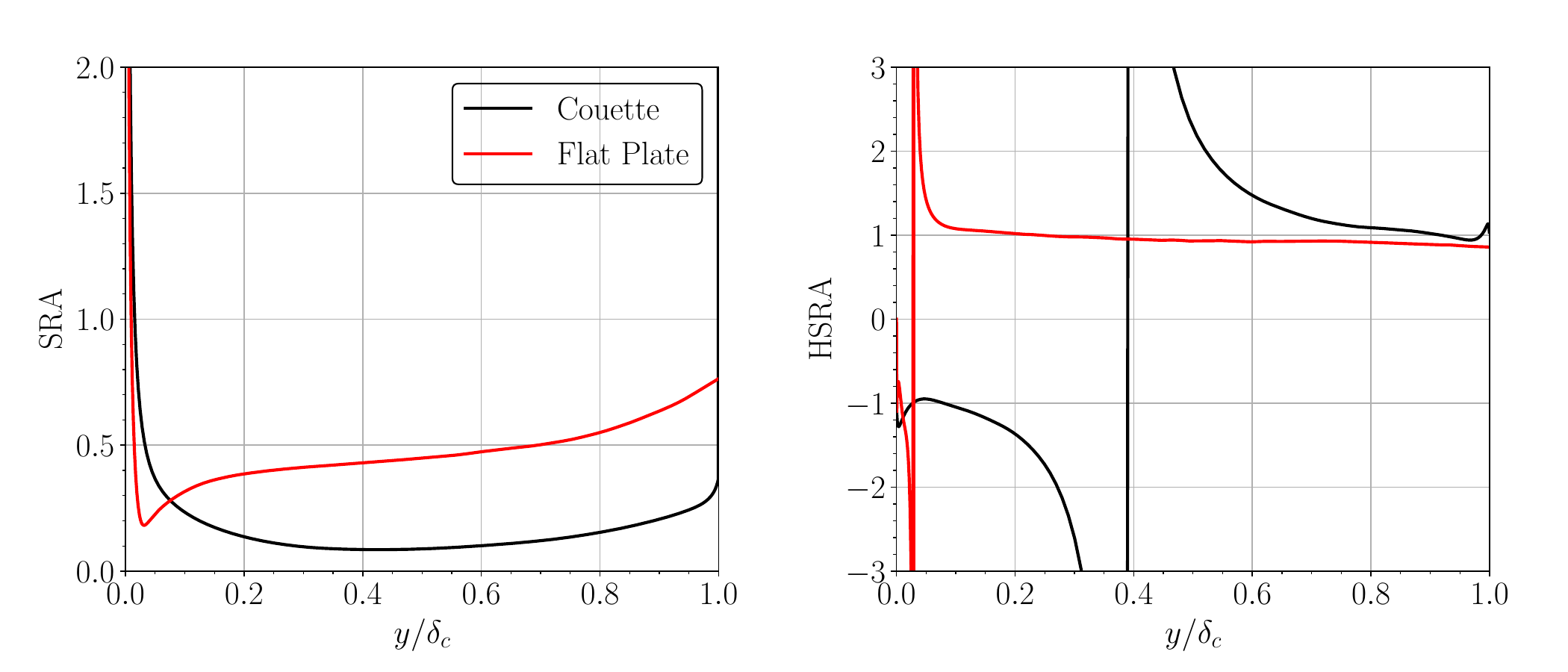}
\put(-370,157){(a)}
\put(-180,157){(b)}
\caption{(a) SRA and (b) HSRA as a function of outer scaling $y/\delta_c$. Here $\delta_c$ is $2\delta$ for the Couette and the boundary-layer thickness $\delta$, for the flat plate.}%
\label{SRA HSRA}%
\end{figure}%

Figure~\ref{SRA HSRA}(a) compares the SRA, given by equation~(\ref{eq:SRA}), for the Couette and flat-plate flows. The SRA is derived under the assumption of negligible total temperature fluctuations, which is not valid for the present configurations because the strongly cooled walls induce significant wall-normal variations in the total temperature. To account for these variations, \citet{huang1995compressible} extended the velocity--temperature fluctuation relationship proposed by \citet{gaviglio1987reynolds} and \citet{rubesin1990extra}, leading to the modified strong Reynolds analogy (HSRA),
\begin{equation}
\frac{\sqrt{\langle T"T"\rangle}/\langle T\rangle}{(\gamma-1)M^2(\sqrt{\langle u"u"\rangle}/\langle u\rangle)} \left(1-\frac{\partial\langle T_0\rangle }{\partial \langle T \rangle}\right) Pr_t\approx 1,
\label{eq:HSRA}
\end{equation}
where $T_0$ is the total temperature. Following \citet{toki_JFM_2025}, the present work employs the Favre-averaged form of the HSRA,
\begin{equation}
\frac{\sqrt{{ T"T"}}/\{ T\}}{(\gamma-1)M^2(\sqrt{{ u"u"}}/\{ u\})} \left(1-\frac{\partial \{ T_0\} }{\partial \{ T \}}\right) Pr_t\approx 1,
\label{eq:HSRA Favre}
\end{equation}
shown in figure~\ref{SRA HSRA}(b) for the Couette and flat-plate flows.

Figure~\ref{SRA HSRA}(a) compares the SRA, given by equation~(\ref{eq:SRA}), for the Couette and flat-plate flows. The SRA deviates significantly from unity over most of the wall-normal extent in both flows. For the Couette flow, the SRA exhibits large values near the walls but decreases substantially toward the channel core. In the flat-plate flow, the SRA increases gradually toward the boundary-layer edge, although it remains noticeably different from unity throughout the domain.

The HSRA provides a significant improvement over the SRA by accounting for the wall-normal variation of the total temperature. In both configurations, the local divergence of the HSRA originates from the turbulent Prandtl number term. For the flat-plate flow, the divergence occurs near the wall, where the mean temperature reaches its maximum value; however, the HSRA remains nearly constant for $y/\delta_c > 0.1$. In contrast, the HSRA for the Couette flow shows negative values away from the wall and exhibits a local divergence around $y/\delta_c = 0.4$, which corresponds to the location of the maximum mean temperature. Beyond this location, the HSRA approaches unity toward the upper wall.

\subsection{Energy spectra \label{sec:spectra}}

\begin{figure}
\centering
\includegraphics[width=1\textwidth]{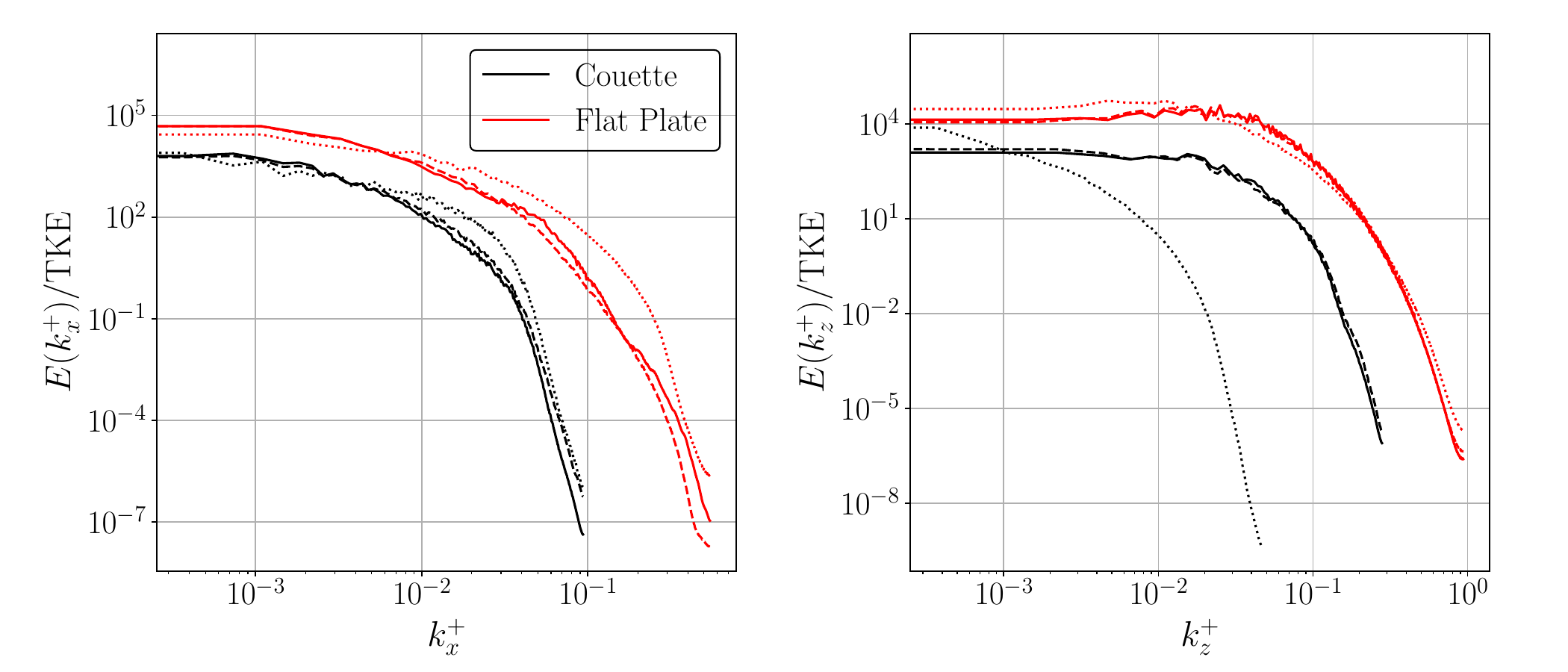}
\put(-370,160){(a)}
\put(-180,160){(b)}
\caption{Normalized density weighted energy spectra in the (a) streamwise and (b) spanwise directions (with turbulent kinetic energy) $E(k_x^+)$ and $E(k_z^+)$ vs ${k_x^+}$ and ${k_z^+}$ respectively at $y^*=5$ (---), $y^*=12$ ($--$) and $y^*=100$ ($\cdot\!\cdot\!\cdot$).}%
\label{density weighted Energy vs k}%
\end{figure}%

The second-order statistics revealed that the turbulent characteristics are similar between the Couette and flat-plate flows below the buffer layer, whereas discrepancies arise above this region. To further investigate these differences, the energy spectra are examined. 

The density weighted energy spectra in the streamwise and spanwise directions for the Couette and flat-plate flows are plotted at three different $y^*$ locations against the non-dimensional wavenumbers $k_x^+$ and $k_z^+$ in figure~\ref{density weighted Energy vs k}. The wavenumbers are non-dimensionalised using the viscous length scale, $\delta_{v}$. In the streamwise direction, a Hanning window is applied to the velocity data for the flat plate to mitigate spectral leakage; without this windowing, the spectra exhibit an artificial plateau at high wavenumbers.

The streamwise energy spectra exhibit good agreement across the three \(y^*\) locations. Compared to the Couette flow, the flat-plate boundary layer displays higher energy levels, with appreciable energy persisting at higher wavenumbers. The spanwise energy spectra for the flat-plate boundary layer similarly show good agreement across the three \(y^*\) locations under the chosen normalization. In contrast, for the Couette flow, the spectrum at \(y^* = 100\) differs from those at \(y^* = 5\) and \(y^* = 12\), exhibiting a narrower range of active scales and reduced energy at higher wavenumbers. This behaviour is consistent with the observations of \citet{yao2023study}, who report the presence of large-scale structures in the core region of Couette flow. These structures concentrate energy at large scales, leading to a rapid decay at higher wavenumbers.

\section{Sub filter scale analysis} \label{SFS analysis}
To compare sub-filter-scale (SFS) phenomena between Couette flow and the flat-plate boundary layer, an explicit spatial filter is applied to the DNS data as a post-processing procedure to extract the SFS contributions. This filtering operation provides an idealised separation of resolved and unresolved scales, analogous to the implicit filtering inherent in LES. 

In \S~\ref{Effect of filter type}, the effect of different filter types, namely the sharp spectral filter and the top-hat filter, is analysed. In \S~\ref{Effect of filter width}, the sharp spectral filter is selected to investigate the effect of filter width. The decomposition of the SFS shear stress, trace of the stress tensor, and wall-normal heat flux is presented in \S~\ref{Decomposition of SFS terms}, where the contributions of the decomposed terms to the overall SFS quantities are examined. To further understand the underlying SFS dynamics, fluctuation fields and SFS quantities are analysed in cross-stream planes, while the SFS quantities and their corresponding residual terms are investigated in wall-parallel planes in \S~\ref{Turbulent Structural Analysis}.

\subsection{Effects of filter type} \label{Effect of filter type}

To assess the sensitivity of the extracted SFS quantities to the filter type, two representative filters are considered: the top-hat filter and the sharp spectral filter. The top-hat filter gradually attenuates small-scale fluctuations, introducing an effective dissipative behaviour, whereas the sharp spectral filter applies an idealised cut-off in spectral space, retaining all scales below the filter width while removing higher-wavenumber components without additional dissipation. This distinction provides an analogue of the role of numerical discretisation in LES simulations. In particular, changing the filter type in the present analysis is analogous to changing the numerical scheme in LES, as different filters introduce different treatments of high-wavenumber content and levels of numerical dissipation.

\begin{figure}
\centering
\includegraphics[width=1\textwidth]{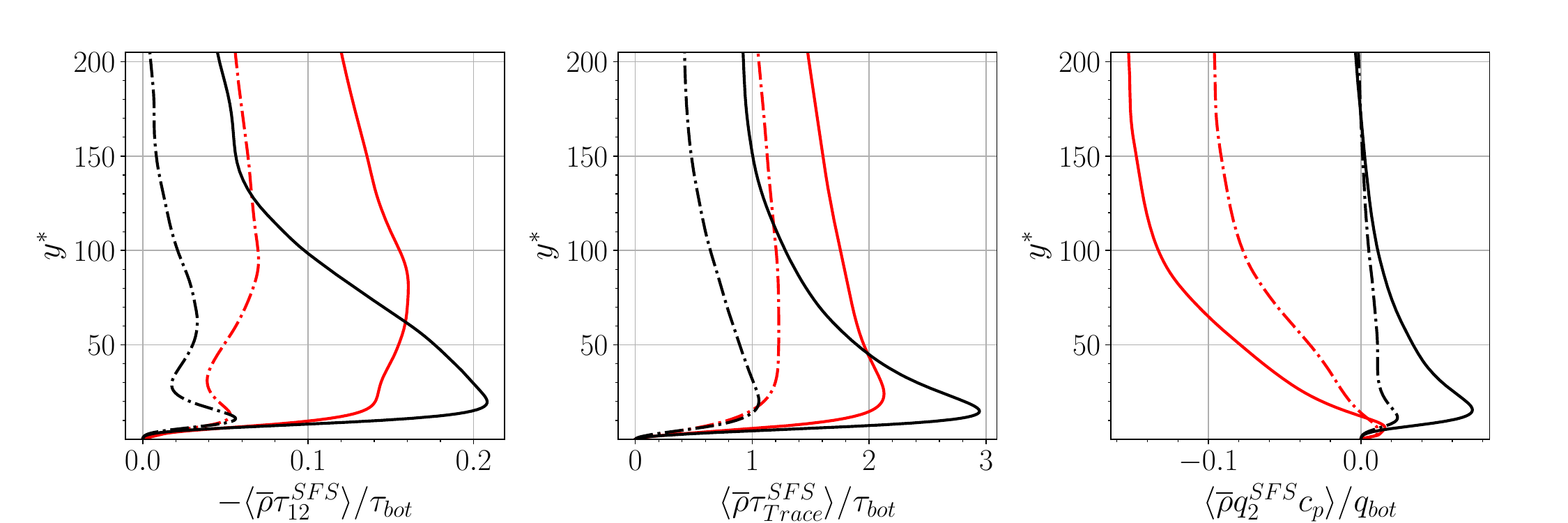}
\put(-370,120){(a)}
\put(-250,120){(b)}
\put(-130,120){(c)}
\caption{Exact sub-filter scale averaged (a) shear stress, (b) trace of stress tensor and (c) heat flux obtained using sharp spectral filter (-$\cdot$-) and top hat filter (---) for filter widths $\overline{\Delta}^{(2)}$ for the hypersonic Couette flow (black) and flat plate (red).}
\label{Stat avg SFS Top Sharp}%
\end{figure}

Figure~\ref{Stat avg SFS Top Sharp} presents the statistically averaged sub-filter-scale (SFS) shear stress, trace of the stress tensor, and heat flux as functions of $y^*$ for both the flat-plate boundary layer and Couette flow using a filter width of $\overline{\Delta}^{(2)}$. Results are shown for both the top-hat and sharp spectral filters. The results show that the top-hat filter yields larger SFS contributions than the sharp spectral filter. This difference arises from the distinct spectral characteristics of the two filters. As shown in figure~\ref{Transfer function}, the top-hat filter exhibits a gradual attenuation of high-wavenumber components, whereas the sharp spectral filter applies an abrupt cut-off. As a result, for the same filter width, the top-hat filter attenuates a broader range of wavenumbers, including part of the lower-wavenumber content, leading to larger SFS magnitudes.

The SFS shear stress profile in figure \ref{Stat avg SFS Top Sharp} (a) exhibits a negative peak around $y^* \approx 30$ for both the flat-plate boundary layer and Couette flow when a sharp spectral filter is employed. This deficit was previously reported for Couette flow by \citet{toki2024sub}, where it was shown to arise from an enhancement of Q1 and Q3 motions, indicating a change in the relative contributions of different quadrant events. The present results further demonstrate that a similar deficit is also observed in the flat-plate boundary layer, indicating that this feature is not unique to Couette flow. The SFS shear stress further exhibits a peak--valley--peak structure, with an initial peak at $y^* \approx 10$, followed by the deficit region, and, in the case of Couette flow, a secondary peak at larger $y^*$ locations.

In contrast, when a top-hat filter is applied, the deficit is no longer observed, and the SFS shear stress profile becomes smoother. This behaviour can be attributed to the gradual attenuation of high-wavenumber components inherent in the top-hat filter, which effectively distributes the scale separation over a broader range of wavenumbers. As a result, the SFS quantities contain contributions from a different range of scales compared with the sharp spectral filter. It is therefore plausible that this change in scale separation alters the relative importance of different quadrant events, with the enhanced contributions of Q1 and Q3 motions observed for the sharp spectral filter being reduced, while Q2 and Q4 motions become relatively more prominent when the top-hat filter is applied. These observations indicate that the deficit in the SFS shear stress is sensitive to the choice of filter, highlighting that SFS behaviour depends not only on the underlying flow physics but also on the definition of the resolved and unresolved scales. This dependence has important implications for LES, where the scale separation is influenced by the numerical discretisation.

The SFS trace of the stress tensor in figure \ref{Stat avg SFS Top Sharp} (b) reaches a maximum in the buffer region for $20 \lesssim y^* \lesssim 40$ in both the Couette flow and the flat-plate boundary layer for both filter types. Consistent with the behaviour observed in the SFS shear stress, the top-hat filter yields larger values than the sharp spectral filter in both configurations. For the Couette flow, the SFS trace decreases towards the channel centre following the peak. In contrast, for the flat-plate boundary layer, the SFS trace exhibits a more gradual variation in the wall-normal direction for both filters, with a weaker decay compared to the Couette flow. These differences can be interpreted in terms of the wall-normal variation of thermophysical properties and the corresponding changes in turbulent scales. In Couette flow, the temperature increases towards the channel centre, leading to an increase in viscosity and promoting larger turbulent structures away from the wall. In contrast, in the flat-plate boundary layer, the temperature decreases above the buffer region towards the boundary-layer edge, resulting in lower viscosity and the persistence of smaller turbulent structures in the outer region. As a result, the characteristic turbulent length scales evolve differently in the two configurations, and this difference is reflected in the SFS trace.

The SFS heat flux in figure \ref{Stat avg SFS Top Sharp} (c) exhibits a peak in the near-wall region for the Couette flow, located at $y^* \approx 10$ for the sharp spectral filter and slightly further away from the wall for the top-hat filter. Beyond this peak, the SFS heat flux decreases towards the channel centre. For the flat-plate boundary layer, the SFS heat flux is positive in the near-wall region ($y^* \lesssim 10$) for both filter types. Further away from the wall, the SFS heat flux becomes negative. This behaviour can be interpreted in terms of the wall-normal variation of the mean temperature gradient. In the Couette flow, the mean temperature gradient decreases towards the channel centre and approaches zero, leading to a corresponding reduction in the magnitude of the SFS heat flux.  In contrast, in the flat-plate boundary layer, the mean temperature gradient changes sign within the buffer region. As a result, the SFS heat flux also changes sign, with positive values in the near-wall region and negative values further away from the wall. Moreover, the magnitude of the SFS heat flux increases with the magnitude of the temperature gradient, indicating that the unresolved scales contribute more significantly to heat transport in regions with a stronger thermal gradient.

\subsection{Effects of filter width \label{Effect of filter width}}

\begin{figure}
\centering
\includegraphics[width=1\textwidth]{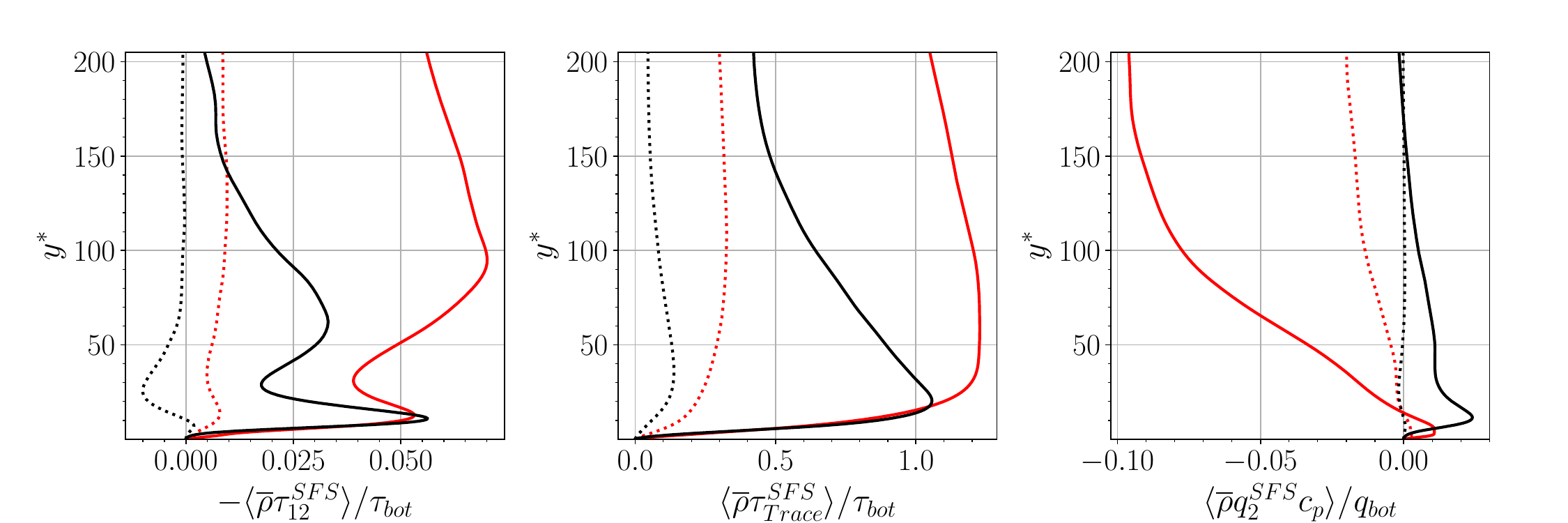}
\put(-370,120){(a)}
\put(-250,120){(b)}
\put(-130,120){(c)}
\caption{Exact sub-filter scale averaged (a) shear stress, (b) trace of stress tensor and (c) and heat flux obtained using sharp spectral filter for $\overline{\Delta}^{(1)}$ ($\cdots$) and $\overline{\Delta}^{(2)}$ (---) for the hypersonic Couette flow (black) and flow over a flat plate (red). }
\label{Stat avg SFS SF flat couette}%
\end{figure}%

In this section, the effect of the filter width on the SFS quantities is examined using the sharp spectral filter. Figure~\ref{Stat avg SFS SF flat couette} compares the SFS quantities for the Couette flow and the flat-plate boundary layer for two filter widths, $\overline{\Delta}^{(1)}$ and $\overline{\Delta}^{(2)}$.

For the Couette flow, increasing the filter width modifies the structure of the SFS shear stress, with the emergence of multiple peaks and a broader distribution across the channel. In contrast, the flat-plate boundary layer exhibits a more gradual variation in the wall-normal direction, with enhanced contributions extending further into the outer region. A similar trend is observed for the SFS stress trace, where increasing the filter width leads to larger amplitudes in both configurations. For the SFS heat flux, the effect of the filter width is particularly pronounced in the Couette flow, where increasing the filter width leads to the emergence of a distinct peak in the buffer region. In the flat-plate boundary layer, the overall shape of the SFS heat flux remains similar for both filter widths, although its magnitude increases with the filter width.

Overall, these results indicate that the SFS quantities are strongly influenced by the filter width, which alters both their magnitude and spatial distribution. This dependence highlights the importance of considering the effective filter width in LES, where the filter scale is closely related to the grid resolution.

\subsection{Decomposition of SFS terms} \label{Decomposition of SFS terms}
In this section, the SFS terms are mathematically decomposed to examine the contributions from different components. A decomposition of the SFS shear stress in hypersonic Couette flow was previously performed by \citet{toki2024sub}. The present analysis extends this approach to both the SFS shear stress and heat flux, and applies it to both Couette and flat-plate configurations.

Following \citet{toki2024sub}, the decomposition of the SFS shear stress is given by
\begin{equation}
 - \langle \overline{\rho} \tau ^\text{SFS} _{12} \rangle
 = -\langle \rho ^{\prime} \widetilde{u} \widetilde{v} \rangle -\langle \rho  u^{\prime \prime} \widetilde{v} \rangle - \langle \rho  \widetilde{u} v^{\prime \prime} \rangle - \langle \overline{\rho} u^{\prime \prime} v^{\prime \prime} \rangle - \langle \rho ^{\prime} u^{\prime \prime} v^{\prime \prime} \rangle.
 \label{decomp_final}
\end{equation}
Applying the same decomposition to the SFS heat flux yields
\begin{equation}
  \langle \overline{\rho} q ^\text{SFS}_{2} \rangle
 = \langle \rho ^{\prime} \widetilde{v} \widetilde{T} \rangle + \langle \rho  v^{\prime \prime} \widetilde{T} \rangle + \langle \rho  \widetilde{v} T^{\prime \prime} \rangle + \langle \overline{\rho} v^{\prime \prime} T^{\prime \prime} \rangle + \langle \rho ^{\prime} v^{\prime \prime} T^{\prime \prime} \rangle.
 \label{decomp_final SFS heat}
\end{equation}
Similarly, summing the diagonal components of the SFS stress tensor gives the trace
\begin{multline}
    \langle \overline{\rho} \tau ^\text{SFS}_\text{Trace} \rangle
 = \langle \rho ^{\prime} (\widetilde{u}^2 + \widetilde{v}^2 + \widetilde{w}^2) \rangle 
 + 2\langle \rho  (u^{\prime \prime} \widetilde{u} + v^{\prime \prime} \widetilde{v} + w^{\prime \prime} \widetilde{w}) \rangle \\
    + \langle \overline{\rho} (u^{\prime \prime 2} + v^{\prime \prime 2} + w^{\prime \prime 2}) \rangle 
    + \langle \rho ^{\prime} (u^{\prime \prime 2} + v^{\prime \prime 2} + w^{\prime \prime 2}) \rangle.
    \label{decomp_final SFS stress trace}
\end{multline}

In equations (\ref{decomp_final}) and (\ref{decomp_final SFS heat}), the first three terms represent interactions between the resolved and residual fields. For the stress trace in equation (\ref{decomp_final SFS stress trace}), these interaction contributions are contained in the first two terms. The subsequent terms correspond to correlations involving the residual components, including second-order and third-order correlations with density and velocity or temperature fluctuations.

\begin{figure}
\centering
\includegraphics[width=1\textwidth]{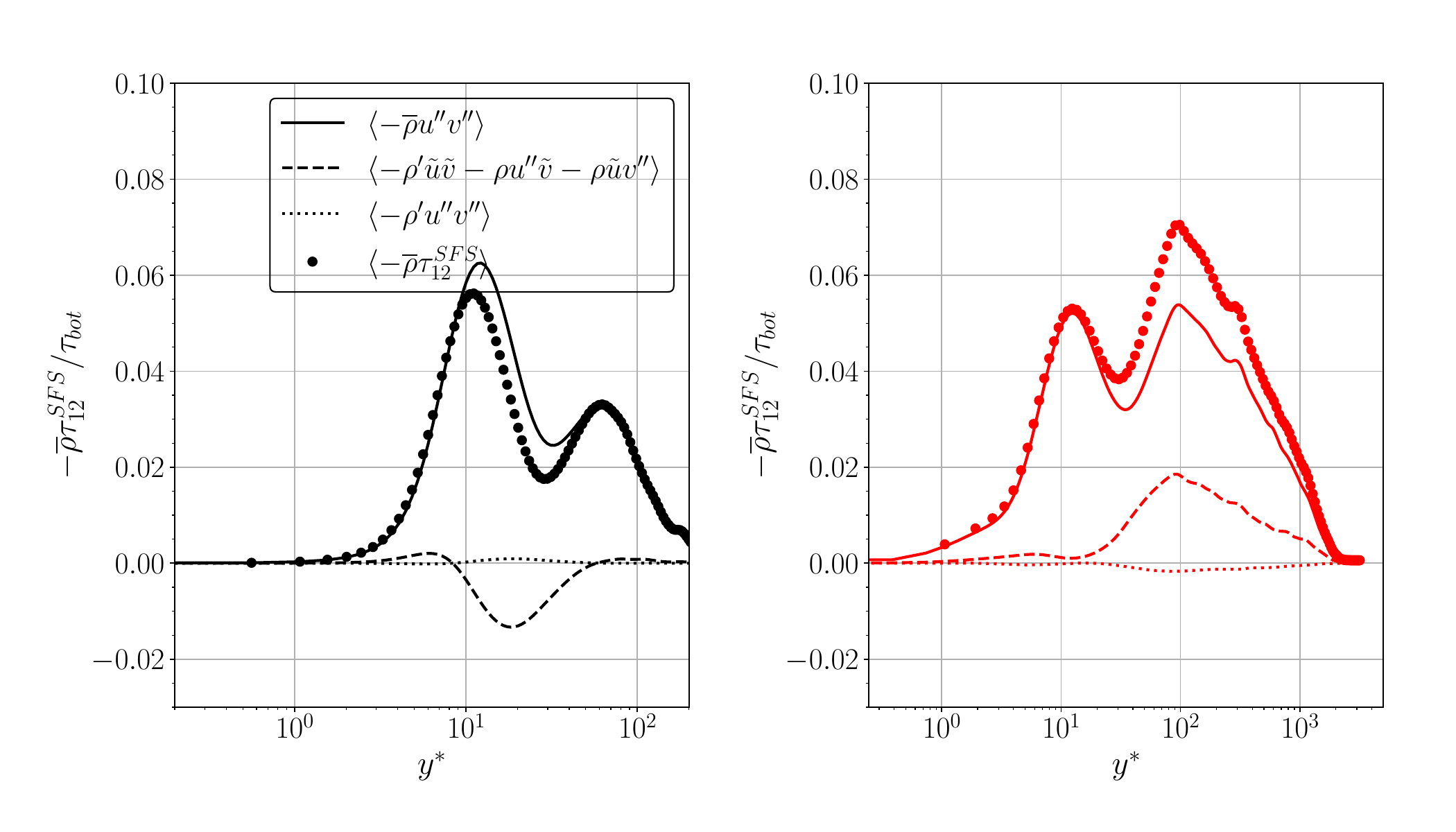}
\put(-370,210){(a)}
\put(-180,210){(b)}
\caption{Mean profiles for the SFS shear stress terms as a function of $y^*$ for the (a) Couette flow and (b) flat plate.  }
\label{Residual shear stress couette flat plate mean}%
\end{figure}%
\begin{figure}
\centering
\includegraphics[width=1\textwidth]{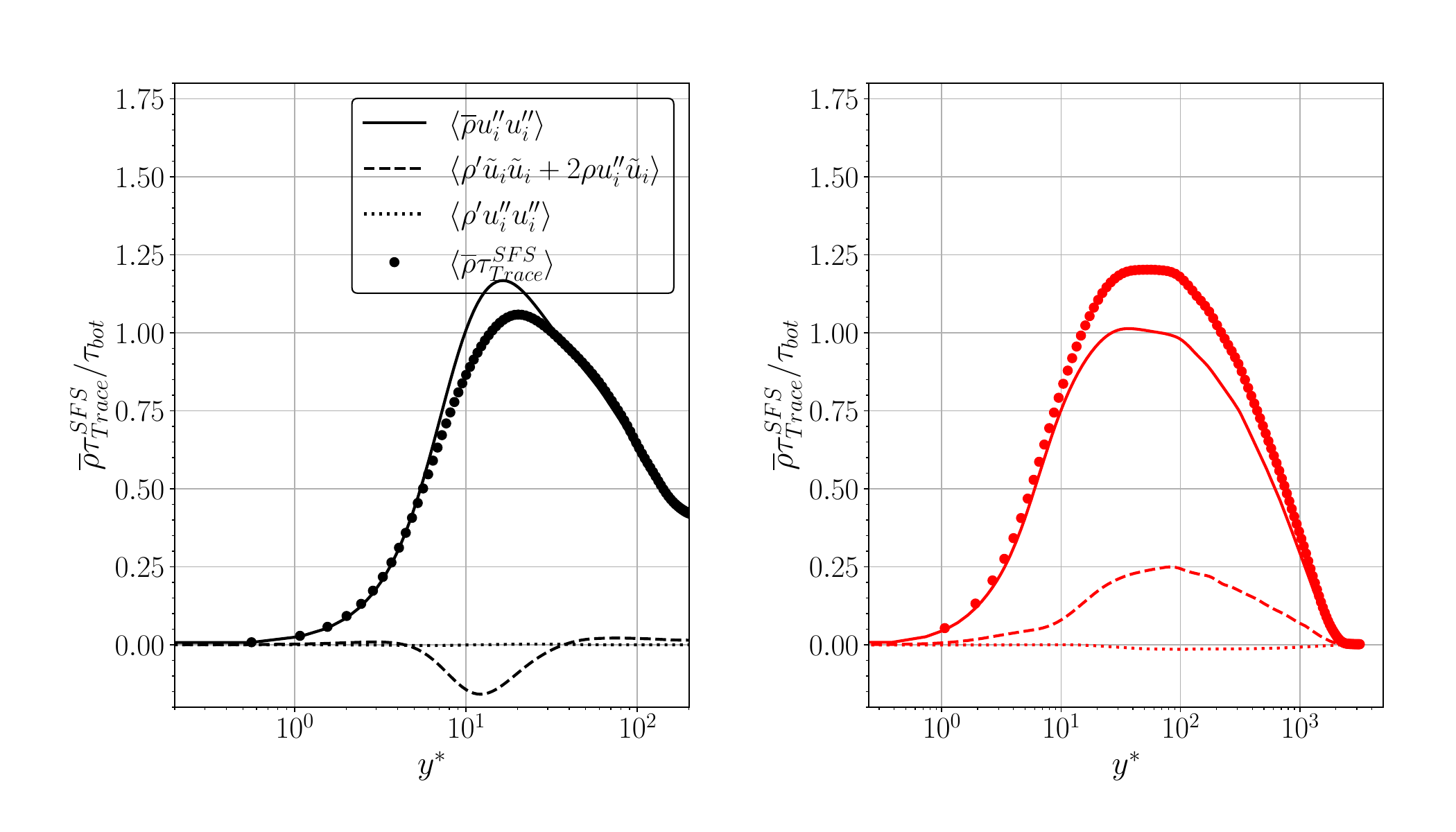}
\put(-370,200){(a)}
\put(-180,200){(b)}
\caption{Mean profiles for the SFS trace terms as a function of $y^*$ for the (a) Couette flow and flat (b) plate. }
\label{Residual Stress Trace couette flat plate mean}%
\end{figure}%
\begin{figure}
\centering
\includegraphics[width=1\textwidth]{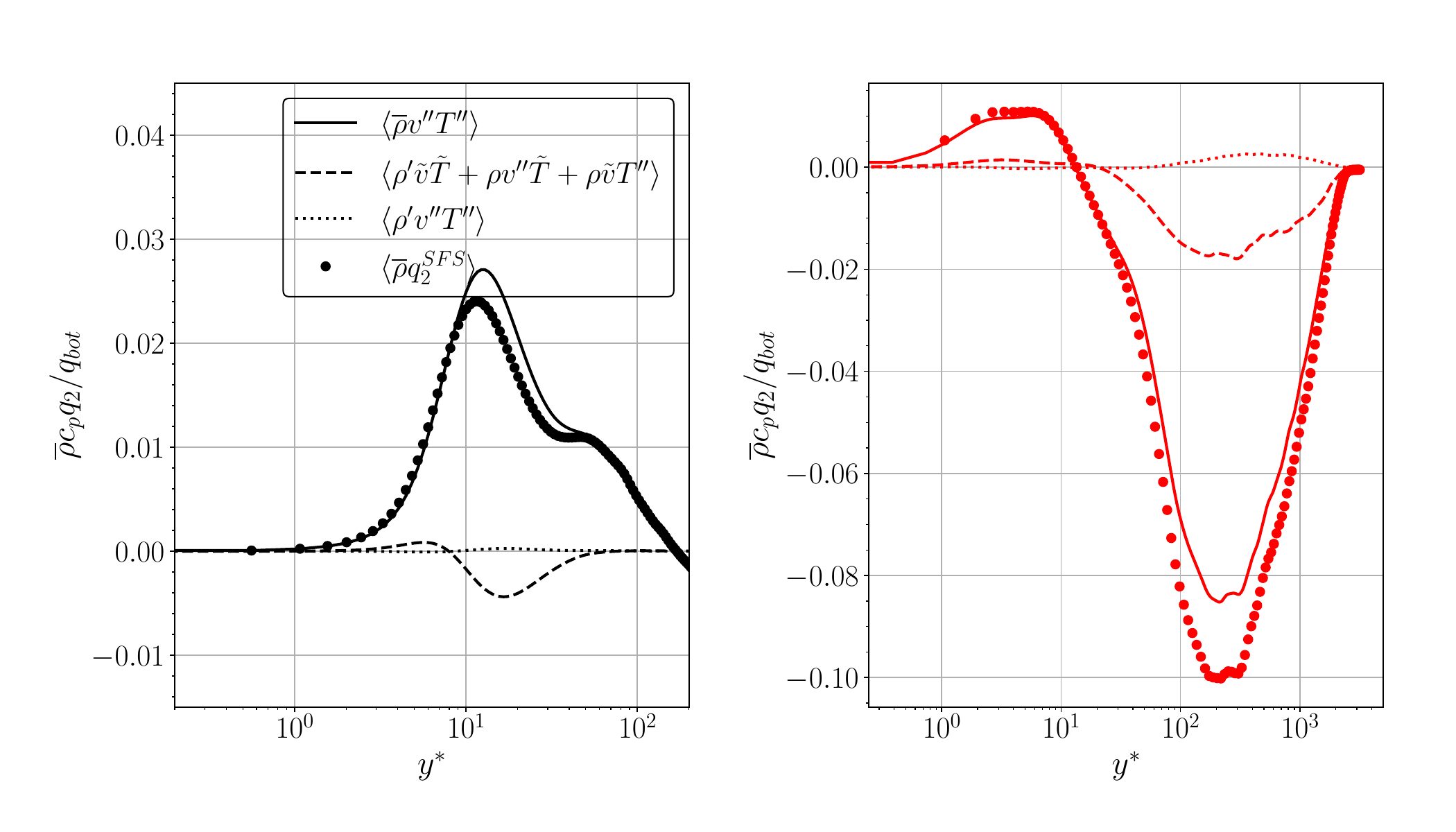}
\put(-370,200){(a)}
\put(-180,200){(b)}
\caption{Mean profiles for the SFS heat flux terms as a function of $y^*$ for the (a) Couette flow and (b) flat plate.}
\label{Residual Heat flux couette flat plate mean}%
\end{figure}%

Figure~\ref{Residual shear stress couette flat plate mean} presents the decomposition of the SFS shear stress into its constituent terms for both Couette flow and the flat-plate boundary layer. The two cases exhibit similar relative contributions from the decomposed terms. In both flows, the second-order residual-velocity correlation provides the dominant contribution to the SFS shear stress, with a peak at $y^* \approx 12$. This indicates that the SFS shear stress is primarily governed by correlations between sub-filter-scale velocity fluctuations. In contrast, the third-order correlation term contributes negligibly, suggesting that higher-order density–velocity interactions play a minor role in the overall SFS stress. Up to $y^* \approx 12$, the SFS shear stress is largely determined by the residual-velocity term in both configurations. Beyond this location, differences between the two flows become apparent. In the Couette flow, the interactions between the resolved and residual fields acts to reduce the total SFS shear stress, whereas in the flat-plate boundary layer it enhances it.

The dominance of the residual-velocity term indicates that the SFS shear stress is governed primarily by turbulent motions. In the Couette flow, the SFS deficit has been shown by \citet{toki2024sub} to arise from an enhancement of Q1 and Q3 motions. The present results indicate that similar behaviour is observed in the flat-plate boundary layer, suggesting that the same turbulent mechanism is active in both configurations. The difference in the sign of the residual contributions between the two flows can be attributed to density variations induced by temperature gradients. In the Couette flow, the temperature increases away from the wall, whereas in the flat-plate boundary layer, the temperature decreases above the buffer region. As a result, the density-related contributions to the SFS shear stress change sign, leading to opposite effects on the total SFS stress in the two configurations.

Figure~\ref{Residual Stress Trace couette flat plate mean} presents a comparison of the decomposed terms in equation~(\ref{decomp_final SFS stress trace}) for the trace of the SFS stress tensor for both Couette flow and the flat-plate boundary layer. The two cases exhibit similar relative contributions from the decomposed terms.

For the Couette case, the residual (second-order) term again provides the dominant contribution to the SFS stress trace, while the third-order correlation term contributes negligibly. In the Couette flow, the SFS stress trace within the viscous sublayer is almost entirely dominated by the residual term. This indicates that the small-scale energy is primarily governed by sub-filter-scale motions.

In contrast, for the flat-plate boundary layer, although the residual term remains the primary contributor, a small but non-negligible contribution from the resolved-flow term persists within the viscous sublayer and near the boundary-layer edge. Furthermore, the resolved-flow contribution is negative for Couette flow, thereby reducing the total SFS stress trace, whereas it is positive for the flat-plate case, leading to an increase in the total SFS stress trace. This difference arises from the opposite signs of the interactions between the resolved and residual fields in the two flows, which reduces the SFS stress trace in the Couette flow and increases it in the flat-plate boundary layer.

Figure~\ref{Residual Heat flux couette flat plate mean} presents the decomposition of the sub-filter-scale (SFS) heat flux for both Couette flow and the flat-plate boundary layer. For both flows, the residual term provides the dominant contribution to the SFS heat flux, while the third-order correlation term is negligible.

For the Couette flow, the profiles of the decomposed SFS heat-flux terms closely resemble those of the decomposed SFS shear stress shown in Figure~\ref{Residual shear stress couette flat plate mean}, except for the absence of the secondary peak observed in the SFS shear stress and its corresponding residual contribution.

In contrast, the decomposed SFS heat-flux terms for the flat-plate boundary layer differ markedly from those of the SFS shear stress. Unlike the SFS shear stress, whose behaviour is primarily associated with turbulent momentum transport, the SFS heat flux represents the transport of thermal energy by sub-filter-scale turbulence and is strongly influenced by the wall-normal temperature distribution. The difference between the two flows therefore reflects their distinct thermal structures.

From the decomposed profiles of $\tau_{12}^{\text{SFS}}$, $\tau_{\text{Trace}}^{\text{SFS}}$, and $q_2^{\text{SFS}}$, the residual component is the dominant contribution across all three quantities. The profiles and magnitudes of these residual terms closely match those of the corresponding SFS quantities, reflecting the dominant contribution of the residual component.

\subsection{Flow structures underlying the SFS behaviour} \label{Turbulent Structural Analysis}

\begin{figure}
\centering
\includegraphics[width=1\textwidth]{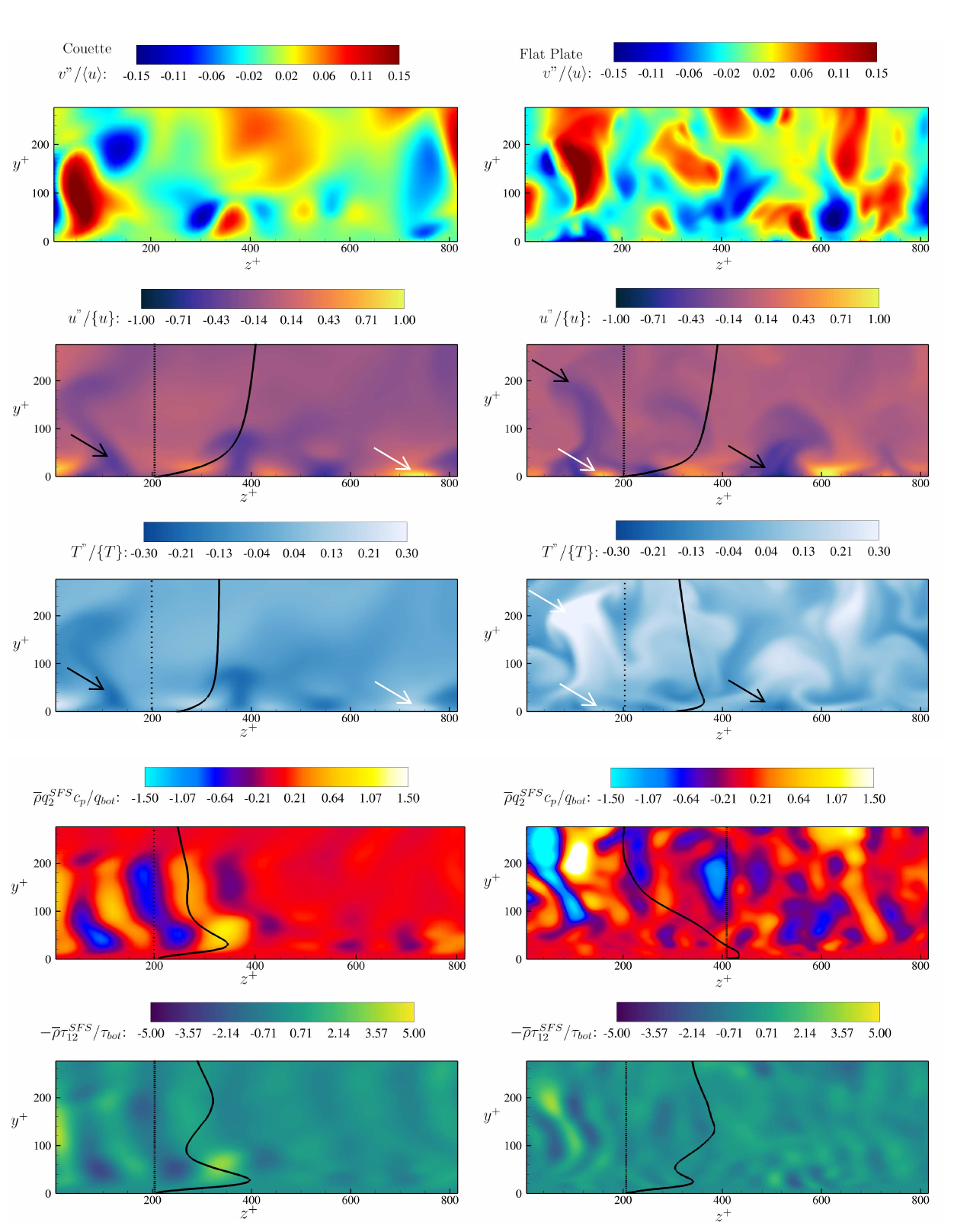}
\put(-190,480){(a)}
\put(-190,380){(b)}
\put(-190,280){(c)}
\put(-190,185){(d)}
\put(-190,85){(e)}
\caption{
Instantaneous snapshots in the cross-stream plane of the normalized 
(a) wall-normal velocity fluctuation, 
(b) streamwise velocity fluctuation, 
(c) temperature fluctuation, 
(d) sub-filter-scale (SFS) wall-normal heat flux, and 
(e) SFS shear stress. 
Each panel compares the Couette flow (left) and flat-plate flow (right). 
In subfigures (b) and (c), black arrows indicate regions of low-velocity or low-temperature fluid, while white arrows indicate regions of high-velocity or high-temperature fluid.}
\label{fluctuation xz}
\end{figure}

The decomposition of the SFS quantities indicates that the contributions to the momentum and heat transport are primarily governed by the residual component. To provide further insight into this behaviour, the underlying turbulent structures are examined in this section.

To examine the wall-normal variation of the turbulent flow field, figure~\ref{fluctuation xz} presents cross-stream ($z^+ - y^+$) planes of the velocity and temperature fluctuations ($u\text{''}$, $v\text{''}$, and $T\text{''}$), together with the SFS wall-normal heat flux, $q_2^{\text{SFS}}$, and the SFS shear stress, $\tau_{12}^{\text{SFS}}$, for both the Couette and flat-plate configurations. The spatial coordinates are expressed using viscous scaling, with $z^+ = \langle \rho_{\text{bot}} \rangle u_{\tau}^+ z / \langle \mu_{\text{bot}} \rangle$ and $y^+ = \langle \rho_{\text{bot}} \rangle u_{\tau}^+ y / \langle \mu_{\text{bot}} \rangle$, where $u_\tau^+ = \sqrt{\tau_{\text{bot}} / \langle \rho_{\text{bot}} \rangle}$. Favre-averaged mean profiles are overlaid to indicate the corresponding wall-normal variation. For the streamwise velocity fluctuations, black arrows indicate low-speed regions, while white arrows indicate high-speed regions. Similarly, for the temperature fluctuations, black arrows indicate low-temperature regions and white arrows indicate high-temperature regions.

The $v\text{''}$ fluctuation contours in figure~\ref{fluctuation xz} (a) exhibit alternating regions of positive and negative $v\text{''}$, corresponding to wall-normal ejection ($v\text{''} > 0$) and sweep ($v\text{''} < 0$) events. The $u\text{''}$ fluctuation contours in figure~\ref{fluctuation xz} (b) show that regions of positive $v\text{''}$ are typically associated with negative $u\text{''}$, and vice versa, in both the Couette and flat-plate flows. This behaviour reflects the well-known anti-correlation between $u\text{''}$ and $v\text{''}$, consistent with the sign and distribution of the Reynolds shear stress.

The temperature fluctuation contours in figure~\ref{fluctuation xz} (c) resemble those of the $u\text{''}$ fluctuations, with low- and high-speed streaks corresponding to regions of low and high temperature. In the Couette flow, this relationship is consistent across the wall-normal direction, reflecting the monotonic increase of the mean temperature profile. In contrast, the flat-plate boundary layer exhibits a distinct double-layer structure. Near the wall, where the mean temperature increases with $y^+$, low-speed streaks correspond to low-temperature regions and high-speed streaks to high-temperature regions. Above the buffer layer, where the mean temperature profile decreases, this correspondence reverses, consistent with the change in the sign of the temperature gradient. The transition between these regions reflects the change in the relationship between velocity and temperature fluctuations caused by the reversal of the mean temperature gradient.

The SFS heat flux and SFS shear stress in figure~\ref{fluctuation xz} (d) and (e) exhibit alternating regions of positive and negative values along the spanwise ($z^+$) direction. These spanwise-alternating structures are more pronounced in the Couette flow than in the flat-plate boundary layer. In the Couette flow, regions of strong SFS activity are concentrated near $y^+ \approx 25$, where both the SFS shear stress and heat flux attain their maximum mean values. This correspondence indicates that the instantaneous SFS structures are concentrated in regions where the mean SFS quantities are large. A secondary region of enhanced activity is also observed around $y^+ \approx 200$, consistent with the secondary peak in the mean SFS shear stress profile, although with weaker magnitude.

In contrast, the flat-plate boundary layer exhibits a different behaviour. The SFS heat flux transitions from positive values near the wall to strongly negative values beyond $y^+ \approx 27$, which is reflected in the instantaneous contours by the dominance of negative regions interspersed with smaller positive patches. The mean SFS shear stress displays two peaks at approximately $y^+ \approx 20$ and $y^+ \approx 130$, and the corresponding contour plots show enhanced spanwise-alternating structures around these locations.

In contrast, the flat-plate boundary layer exhibits a different behaviour. The mean SFS heat flux changes sign in the wall-normal direction, becoming negative beyond $y^+ \approx 27$. The instantaneous contours show a similar trend, with predominantly negative regions in the outer part of the boundary layer and smaller positive regions near the wall. The mean SFS shear stress exhibits two peaks at approximately $y^+ \approx 20$ and $y^+ \approx 130$, and enhanced spanwise-alternating structures are observed around these locations in the instantaneous fields.

Overall, the instantaneous SFS structures in both flows reflect the wall-normal variation of the mean quantities, while exhibiting distinct differences between the Couette and flat-plate configurations.

\begin{figure}
\centering
\includegraphics[width=1\textwidth]{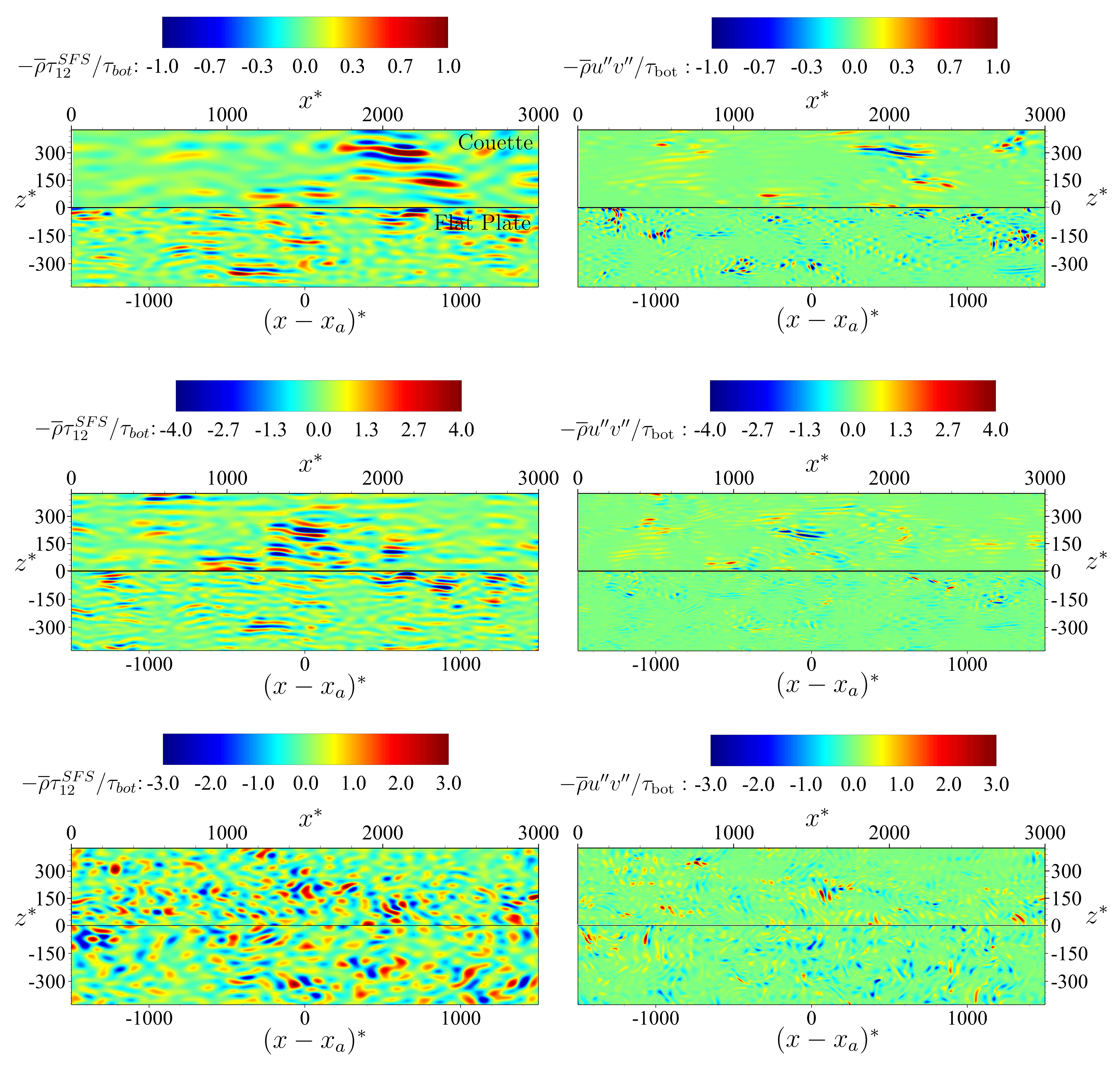}
\put(-370,355){(a)}
\put(-190,355){(d)}
\put(-370,230){(b)}
\put(-190,230){(e)}
\put(-370,110){(c)}
\put(-190,110){(f)}
\caption{
Instantaneous exact SFS normalised shear stress, $-\overline{\rho}\tau_{12}^{\text{SFS}}/\tau_{\text{bot}}$, and residual shear stress, $-\overline{\rho} u^{\prime\prime} v^{\prime\prime}$, for the hypersonic Couette and flat plate flows, obtained using a filter width of $\overline{\Delta}^{(2)}$. Subfigures (a–c) show the exact SFS shear stress at wall-normal locations: (a) $y^* = 5$ , (b) $y^* = 12$, and (c) $y^* = 100$. Subfigures (d–f) show the corresponding residual shear stress at the same wall-normal locations: (d) $y^* = 5$, (e) $y^* = 12$, and (f) $y^* = 100$. Each subfigure compares both flow configurations at the specified height. }

\label{Exact SFS tau12}%
\end{figure}%

\begin{figure}
\centering
\includegraphics[width=1\textwidth]{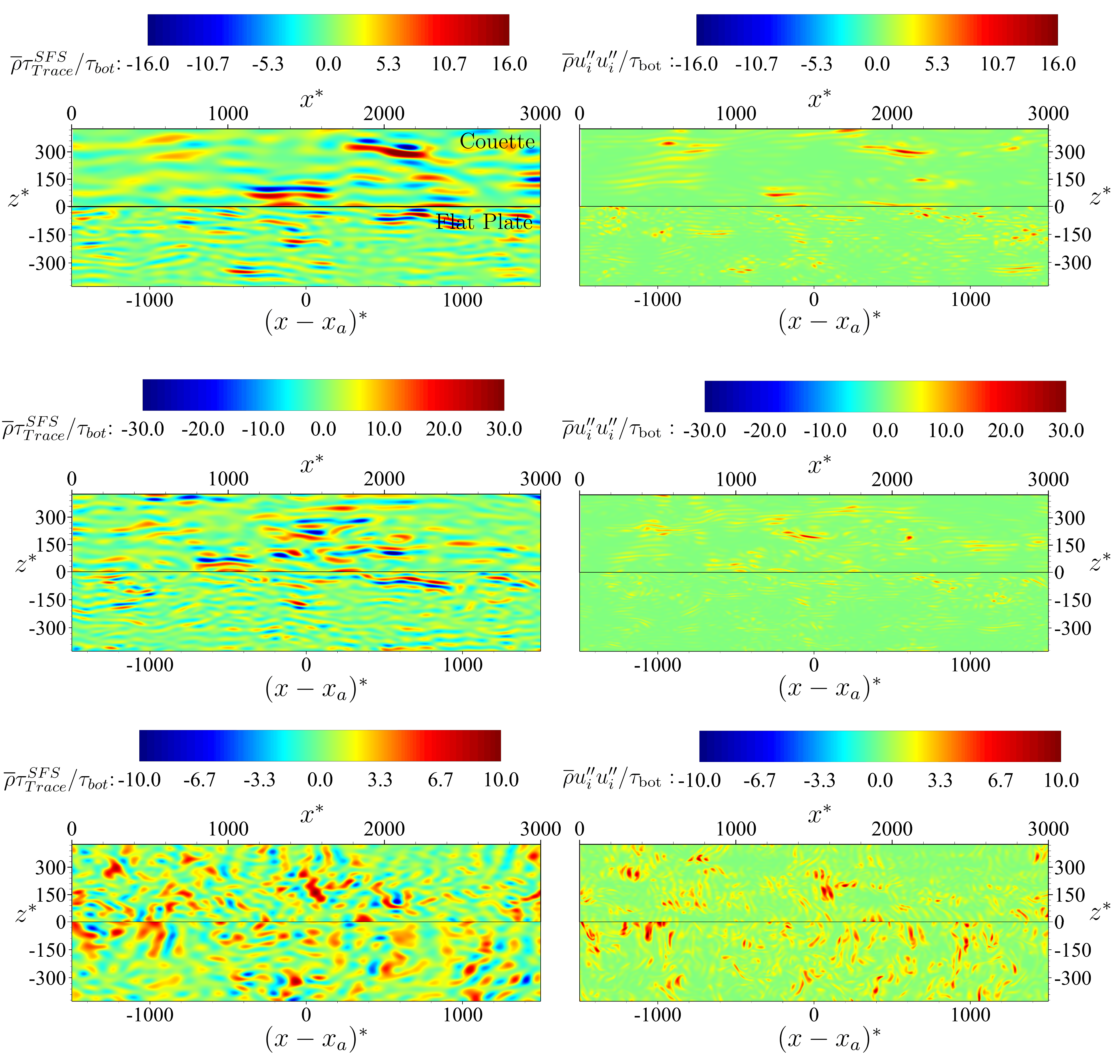}
\put(-370,355){(a)}
\put(-190,355){(d)}
\put(-370,230){(b)}
\put(-190,230){(e)}
\put(-370,110){(c)}
\put(-190,110){(f)}
\caption{
Instantaneous exact sub-filter scale (SFS) normalised stress trace, $\overline{\rho}\tau_{\text{Trace}}^{\text{SFS}} / \tau_{\text{bot}}$, and residual stress trace, $\overline{\rho} u_i^{\prime\prime} u_i^{\prime\prime}$, for the hypersonic Couette and flat plate flows, obtained using a filter width of $\overline{\Delta}^{(2)}$. Subfigures (a–c) show the exact SFS stress trace at wall-normal locations: (a) $y^* = 5$, (b) $y^* = 12$, and (c) $y^* = 100$. Subfigures (d–f) show the corresponding residual stress trace at the same wall-normal locations: (d) $y^* = 5$, (e) $y^* = 12$, and (f) $y^* = 100$. }
\label{Exact SFS tau trace}%
\end{figure}%

\begin{figure}
\centering
\includegraphics[width=1\textwidth]{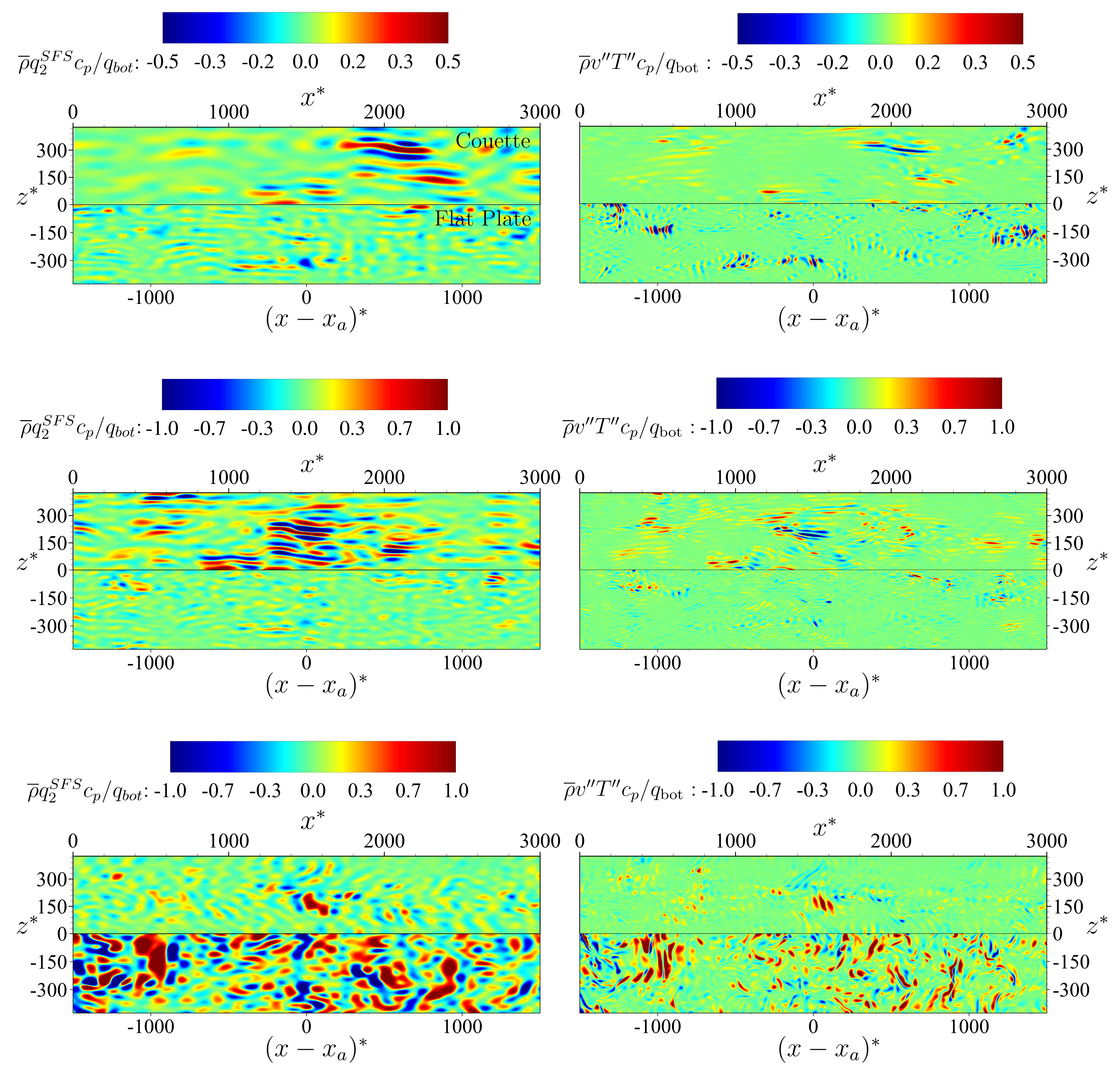}
\put(-370,355){(a)}
\put(-190,355){(d)}
\put(-370,230){(b)}
\put(-190,230){(e)}
\put(-370,110){(c)}
\put(-190,110){(f)}
\caption{
Instantaneous exact sub-filter scale (SFS) normalised wall-normal heat flux, 
$-\overline{\rho}\,q_{2}^{\mathrm{SFS}}\,c_p / q_{\mathrm{bot}}$, 
and residual heat flux, $-\overline{\rho}\,v''\,T''$, for the hypersonic Couette and flat-plate flows, obtained with a filter width of $\overline{\Delta}^{(2)}$. 
Subfigures (a–c) show the exact SFS heat flux at wall-normal locations: 
(a) $y^* = 5$, 
(b) $y^* = 12$, 
and (c) $y^* = 100$. 
Subfigures (d–f) show the corresponding residual heat flux at the same wall-normal locations: 
(d) $y^* = 5$, 
(e) $y^* = 12$, 
and (f) $y^* = 100$. 
}
\label{Exact SFS q2}%
\end{figure}%

Given the dominant role of the residual component in all SFS quantities, the instantaneous fields of the SFS quantities and their corresponding residual terms are examined. These fields are analysed at three representative $y^*$ locations spanning the viscous sublayer, buffer layer, and outer layer.

Figure~\ref{Exact SFS tau12} presents the contours of the SFS shear stress and the residual shear stress for both the Couette and flat-plate boundary layer at three representative $y^*$ locations. The $x$- and $z$-coordinates are scaled using the semi-local scaling for comparison between the two flows. Across all wall-normal locations, the magnitude of the instantaneous SFS shear stress is generally larger than that of the residual shear stress. This indicates that, although the residual term dominates the mean SFS shear stress, the local extrema of the SFS field are strongly influenced by contributions from the remaining decomposed terms. These contributions exhibit both positive and negative fluctuations, which largely cancel under averaging, resulting in a small net contribution to the mean momentum transport.

At $y^* = 5$, the SFS shear stress exhibits elongated streak-like structures, consistent with the characteristic streak structures of near-wall turbulence. These structures become shorter at $y^* = 12$, and are no longer apparent at $y^* = 100$, indicating a transition from anisotropic near-wall structures to more isotropic behaviour in the outer region. The residual shear stress, in contrast, exhibits only weak streak-like features in the Couette flow at $y^* = 5$ and $y^* = 12$, and no clear elongated structures in the flat-plate boundary layer at any wall-normal location. This suggests that, although the residual component dominates the mean SFS shear stress, the coherent streak structures observed in the instantaneous SFS field arise from contributions beyond the residual term.

Figure~\ref{Exact SFS tau trace} presents the instantaneous fields of the normalised $\tau_{\text{Trace}}^{\text{SFS}}$ and its residual component for both the Couette and flat-plate boundary layer at three representative $y^*$ locations. The SFS trace exhibits a similar wall-normal evolution to that observed for the SFS shear stress.

A key distinction from the SFS shear stress lies in the interpretation of the trace as an energy-like quantity. The residual component remains strictly positive at all wall-normal locations, as it represents the sum of squared sub-filter-scale velocity fluctuations. In contrast, the SFS trace exhibits local negative values, indicating that it is not a purely positive-definite energy.

The SFS trace represents an energy-like quantity associated with sub-filter-scale motions, but also includes interactions between resolved and sub-filter-scale components. When these interaction terms become sufficiently large and negative, they can outweigh the positive residual contribution, leading to local negative values of the SFS trace.

The separation between resolved and sub-filter-scale components is introduced by the filtering operation and does not correspond to a strict physical separation of turbulent scales. Therefore, the relative contributions of the two components can vary locally as turbulent structures evolve. The negative SFS trace observed in the present study represents a local redistribution of energy between the resolved and sub-filter scales, rather than the existence of a negative energy component. This behaviour is a general consequence of filter-based scale separation and is not specific to compressible or hypersonic turbulence.

Figure~\ref{Exact SFS q2} presents the instantaneous fields of the SFS wall-normal heat flux, $q_2^{\text{SFS}}$, and its residual component for both the Couette flow and the flat-plate boundary layer at three representative $y^*$ locations.

In the Couette flow, elongated streak-like structures are observed in both the SFS heat flux and the residual field at $y^* = 5$ and $y^* = 12$. These structures resemble near-wall streaks, indicating a degree of structural similarity between the two fields in the near-wall and buffer regions. At $y^* = 100$, such streak-like structures are no longer observed, indicating a transition to more isotropic behaviour.

In contrast, the flat-plate boundary layer does not exhibit clear streak-like structures in either the SFS heat flux or the residual field at any wall-normal location. This suggests that, at the sub-filter scale, the correlation between velocity and temperature fluctuations does not retain coherent streak-like structures.

In terms of magnitude, both the SFS heat flux and the residual component in the flat-plate boundary layer exhibit higher values at $y^* = 100$ than at $y^* = 12$. This trend is consistent with the variation of the mean temperature gradient, which becomes stronger above the buffer region following the change in its sign. This behaviour contrasts with that observed for the SFS shear stress and stress trace, where the largest contributions are typically found closer to the wall. The present results indicate that the SFS heat flux is influenced by the wall-normal variation of the temperature field, leading to enhanced activity away from the near-wall region.

\section{Conclusion}\label{Conclusion}
A comparative analysis of sub-filter-scale (SFS) dynamics between hypersonic Couette flow and a flat-plate boundary layer was conducted using filtered DNS data. The Couette flow features a top wall moving at a Mach number of 8, while the flat plate case has a freestream Mach number of 5.84. The friction Reynolds number for both flows is 713. The wall-to-recovery temperature ratio is 0.112 for the Couette case and 0.25 for the flat plate.

The mean flow and second-order statistics reveal that, although the velocity fields in the Couette flow and flat-plate boundary layer exhibit similar behaviour in the near-wall region, significant differences arise in the log-layer and in the thermal structure of the two flows. The streamwise velocity profiles, velocity fluctuations, and turbulence characteristics show qualitative similarity below the buffer layer for both flows, whereas their outer-layer behaviour differs due to the distinct internal and external flow structures. In the Couette flow, turbulence persists throughout the channel, whereas in the flat-plate boundary layer, the turbulence intensity decreases toward the edge of the boundary layer. This behaviour is reflected in the Reynolds shear stress and fluctuation statistics. The temperature and density fields also differ substantially across the wall-normal direction. In the Couette flow, the temperature increases monotonically and becomes nearly uniform in the channel core, leading to weak thermal gradients. In contrast, the flat-plate boundary layer exhibits a strong temperature peak within the buffer layer, followed by a decrease toward the freestream, resulting in sustained thermal gradients away from the wall. These differences in the thermal field also affect the temperature–velocity relation. In the Couette flow, larger deviations are observed in existing correlations, particularly in the channel core where the temperature becomes nearly uniform, whereas the correlations show better agreement in the flat-plate boundary layer. The turbulent Prandtl number remains close to unity for both flows over most of the wall-normal extent, except in regions where the wall-normal turbulent heat flux changes sign. In contrast, the SRA does not collapse between the two flows and deviates significantly from unity. The HSRA provides a noticeable improvement, particularly for the flat plate and for the Couette flow near the upper wall. The energy spectra highlight differences in the distribution of turbulent scales. The energy spectra in the log layer of the flat-plate boundary layer exhibit broader spectral content, with significant energy at higher wavenumbers, indicating the presence of smaller-scale structures. In contrast, the Couette flow shows a narrowing of the spectral range, with energy concentrated at larger scales in the channel core. These results demonstrate that the two flows exhibit fundamentally different spectral characteristics away from the wall.

The analysis of sub-filter-scale (SFS) quantities demonstrates that both the choice of filter and the filter width have a strong influence on the extracted SFS behaviour. The top-hat and sharp spectral filters impose different treatments of the high-wavenumber content owing to their distinct spectral characteristics. This difference is analogous to the role of numerical discretisation in LES, where different numerical schemes provide different effective treatments of unresolved scales. Increasing the filter width modifies both the magnitude and spatial distribution of the SFS quantities, corresponding to the effect of changing the effective grid resolution in LES. These observations suggest that the SFS behaviour in LES is not determined solely by the turbulence field, but is also influenced by the numerical treatment of unresolved scales. Therefore, SFS models should be developed and assessed while considering their dependence on the underlying numerical treatment.

The decomposition of the SFS terms shows that the residual component is the dominant contribution to both the SFS shear stress and heat flux in the two flow configurations, while higher-order terms remain negligible. Differences between the Couette flow and the flat-plate boundary layer arise in the interaction terms between the resolved and residual fields, which act to reduce the total SFS quantities in the Couette flow but enhance them in the flat-plate case.

The visualization of the instantaneous SFS fields highlights differences between the contributions to the mean and instantaneous SFS fields. Although the residual component dominates the mean SFS quantities, the instantaneous SFS fields exhibit larger magnitudes than their corresponding residual components across all wall-normal locations. This indicates that the local extrema of the SFS fields are influenced by contributions from the remaining decomposed terms, which exhibit both positive and negative fluctuations that largely cancel under averaging.

The behaviour of the SFS stress trace further illustrates the nature of these contributions. While the residual component remains strictly positive, as it represents the sum of squared sub-filter-scale velocity fluctuations, the SFS trace itself exhibits local negative values. This reflects the presence of interaction terms between resolved and sub-filter-scale components, which can locally outweigh the positive residual contribution. The filtering operation does not represent a physically sharp separation of scales, and the energy distribution between resolved and sub-filter-scale components changes with the local turbulent structure. Therefore, the SFS trace should not be interpreted as a strictly positive-definite energy. Negative values arise when energy is redistributed from the sub-filter scales to the resolved scales, producing a negative SFS contribution rather than representing a physical loss of energy.

\appendix

\bigskip
\noindent
\textbf{Acknowledgements.} The authors acknowledge the computational support provided by the Rosen Center for Advanced Computing (RCAC) at Purdue University. Microsoft Copilot (M365 Copilot, GPT-5 Chat model) was used for language editing and improving the clarity of the manuscript. All AI-assisted suggestions were reviewed by the authors, who take full responsibility for the final content of the manuscript.

\bigskip
\noindent
\textbf{Funding.} The authors acknowledge support through N00014-21-1-2475 (ONR-MURI, PI: Venkat Raman), and N00014-23-1-2560 (Core ONR program, PI: Carlo Scalo) with Dr. Eric Marineau as Program Manager.

\bigskip
\noindent
\textbf{Declaration of interests.} The authors report no conflict of interest.

\bigskip
\noindent
\textbf{Author ORCIDs.}

Raagvendra Singh \url{https://orcid.org/0009-0003-7430-9502};

Sabbir Ahmed \url{https://orcid.org/0009-0004-4586-1004};

Lian Duan \url{https://orcid.org/0000-0003-0627-6211};

Carlo Scalo \url{https://orcid.org/0000-0002-6774-2207};

Takahiko Toki \url{https://orcid.org/0000-0002-3408-0681};

\section{Effect of Sharp Spectral Filter for Non-periodic Flow}\label{Sharp Spectral effect}

\begin{figure}
\centering
\includegraphics[width=0.65\textwidth]{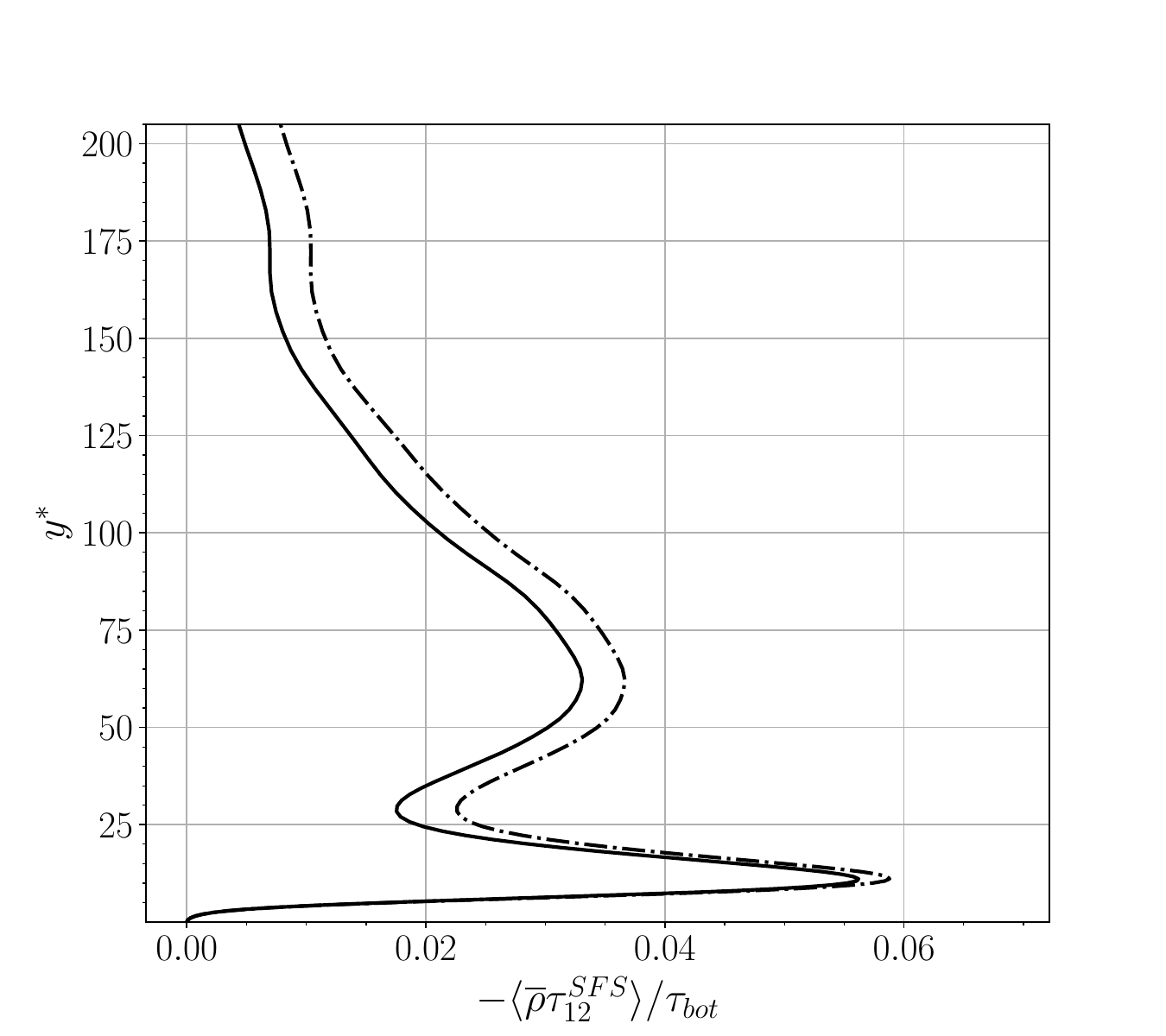}
\caption{Exact sub-filter-scale averaged shear stress for Couette flow obtained using a sharp spectral filter with filter width \(\overline{\Delta}^{(2)}\). The solid line (\(-\)) denotes the SFS shear stress computed by applying the sharp spectral filter to the entire domain, while the dash–dot line (\(-\cdot-\)) denotes the SFS shear stress obtained by averaging the stresses computed separately on the two streamwise halves of the domain.
}
\label{sharp spectral effect fig}%
\end{figure}%

In the present work, a sharp spectral filter is applied in the non-homogeneous direction to obtain sub-filter-scale (SFS) quantities for the hypersonic flat-plate boundary layer. It is well known that applying spectral filters in non-periodic directions can introduce spectral leakage, resulting in the transfer of energy to higher wavenumbers and consequently distorting the filtered field. To assess the impact of applying a sharp spectral filter to non-periodic data, an analysis is conducted using hypersonic Couette flow, which is homogeneous and periodic in the streamwise direction. The computational domain is split into two equal halves, spanning \(0 \leq x \leq 12\delta\) and \(12\delta \leq x \leq 24\delta\). Sub-filter-scale quantities obtained by applying a sharp spectral filter to the full domain are compared with those obtained by applying the same filter separately to the two streamwise halves without any windowing and subsequently averaging the results. Although periodic boundary conditions are imposed on the full domain, the split data series is not periodically continuous. Consequently, each half-domain behaves as a non-periodic signal for the purpose of spectral filtering, allowing an assessment of the error introduced by applying a sharp spectral filter to non-periodic data.

Figure~\ref{sharp spectral effect fig} compares the SFS shear stress obtained using these two approaches. For \(y^* < 10\), the results are in close agreement. However, beyond this location, the decomposed-domain approach yields systematically higher values, with differences reaching approximately \(20\%\) outside the buffer layer. This increase is attributed to spectral leakage introduced by the artificial boundaries, which leads to a redistribution of energy toward higher wavenumbers. As a result, the sharp spectral filter removes a larger fraction of energy, resulting in an overestimation of the SFS shear stress. Despite this discrepancy, the overall trends remain consistent between the two approaches, indicating that the sharp spectral filter applied without windowing captures the qualitative behaviour of the SFS quantities, particularly in the near-wall region. Although this test is performed using Couette flow, it provides an estimate of the error associated with applying a sharp spectral filter in the non-homogeneous streamwise direction. A comparable influence of spectral leakage is expected for the flat-plate boundary layer.

\bibliographystyle{jfm}
\bibliography{ref}

\end{document}